\documentclass[aps,pra,preprint,showpacs,endfloats]{revtex4-1}

\input{Qcircuit}
\usepackage{appendix,amsmath,amssymb,bbm,graphicx,nccmath,subfigure}

\begin{document}

\title{Entanglement dynamics of two-particle quantum walks}
\author{G. R. Carson, T. Loke and J. B. Wang}
\address{School of Physics, The University of Western Australia, WA 6009, Australia}

\begin{abstract}

This paper explores the entanglement dynamics generated by
interacting two-particle quantum walks on degree-regular and -irregular graphs. We
performed spectral analysis of the time-evolution of both the particle probability
distribution and the entanglement between the two particles for various
interaction strength. While the particle probability distributions are stable and not sensitive to perturbations in the interaction strength, the entanglement dynamics are found to be much more sensitive to system variations.  This property may be utilised to probe small differences in the system parameters.

\end{abstract}


\maketitle

\section{Introduction}

Entanglement dynamics has been extensively investigated in the context of
quantum chaos and quantum-classical correspondence. Several nonlinear
models have been examined for their entanglement dynamics, for example, the $ N
$-atom Jaynes-Cummings model \cite{Furuya1999,Angelo1999,Angelo2001}, coupled
kicked tops \cite{Miller1999,Fujisaki2003,Band2004,Kubotani2008}, the Dicke model
\cite{Hou2004}, and Rydberg molecules \cite{Lombardi2006,Lombardi2006b}. Such
studies have focused on the connection between the dynamics of classically chaotic 
systems with the dynamical generation of entanglement in their corresponding quantised systems.

In this work, we study the entanglement dynamics in two-particle discrete-time
quantum walks on degree-regular and -irregular graphs.  Single-particle quantum walks have already emerged as
a useful tool in the development of quantum algorithms \cite{kempe2003a,
Shenvi2003, Reitzner2009, Douglas2008, Berry2010, SmithMosca2012, AnuWang2014, KiaWang2014}. These
algorithms depend on interference between the multiple paths that are
simultaneously traversed by the quantum walker. In multi-particle quantum
walks, the dimension of the state space increases exponentially with the number
of particles and there is the additional resource of interaction and
entanglement between particles, which may be utilised to develop efficient quantum algorithms to solve practically significant applications.

The theoretical study of discrete-time quantum walks with more than one
particle was initiated by Omar \emph{et al.}~\cite{Omar2006}, who considered
non-interacting two-particle quantum walks on the infinite line. Omar \emph{et
al.}~established a role for entanglement in two-particle quantum walks by
showing that initial states which are entangled in their coin degrees of
freedom can generate two-particle probability distributions in which the
positions of the two particles exhibit quantum correlations. Also with
non-interacting particles, \v{S}tefa\v{n}\'{a}k \emph{et
al.}~\cite{Stefanak2006} considered the meeting problem in the discrete-time
quantum walk.

In addition to these studies with distinguishable particles, there have also
been theoretical investigations of two-photon quantum walks 
\cite{Pathak2007,Rohde2011}. By increasing the dimension of the coin Hilbert
space to incorporate both the direction of photon propagation and polarisation,
Pathak and Agarwal \cite{Pathak2007} introduced a quantum walk in which two
photons initially in separable Fock states become entangled through the action
of linear optical elements. While the quantum walk studied in
\cite{Omar2006,Stefanak2006} requires entangled initial states to generate
spatial correlations, the two-photon walk studied in \cite{Pathak2007}, with
its larger coin space, is capable of generating entanglement even from
initially separable states. Venegas-Andraca and Bose \cite{venegas-andracaa}
have also proposed a variant of the two-particle quantum walk in which the
particles have a shared coin space. In this system, entanglement is introduced
between the spatial degrees of freedom of the particles by performing
measurements in the shared coin space.  More recently, Berry and Wang
\cite{Berry2011} introduced two spatial interaction schemes, which have shown
to dynamically generate complex entanglement between the two walking particles.
Rohde et. al. \cite{Rohde2012} studied the entanglement dynamics in a
single-particle quantum walk between its position states on a finite line with
bound and reflecting boundaries. They demonstrated the emergence of
quasi-periodicity in the entanglement time series.

In this work, we explore the detailed entanglement dynamics generated by
locally-interacting two-particle quantum walks on degree-regular and -irregular graphs.  
Our key purpose is to demonstrate that entanglement dynamics arising from two-particle interacting quantum walks on graphs with simple underlying structures can be complex and sensitive to perturbations in the interaction strength parameter.
This work provides strong evidence through sufficiently simple examples where the evolution of the system wavefunction is stable and not sensitive to perturbations to the interaction strength, while the entanglement dynamics of the system is much more sensitive to these small changes. Even in cases where the system wavefunction appears to evolve in a more complex manner, the marginal probability distribution is stable compared to the entanglement between the two particles. Both degree-regular and degree-irregular graphs are studied in this work. Degree-regular graphs are graphs for which every vertex has the same number of neighbours.  We studies three simple degree-regular graphs, including the complete $K_8$ graph (without self-loops), the 3-dimensional hypercube $Q_3$, and the joined 2nd generation 3-Cayley tree. The degree-irregular graphs studied are variations of these degree-regular graphs, with several edges removed.

This paper is structured as follows. In Section \ref{sec:reviewmat} we describe
the mathematical formalism for the interacting two-particle quantum walks and
the evaluation of entanglement between the two particles. In Section
\ref{sec:main}, we perform spectral analysis on the time-evolution of
probability distribution and entanglement time series of two-particle discrete-time quantum walks for
different values of interaction strength. Also presented are the corresponding
Feigenbaum diagrams of the frequencies in the time series. Finally, Section \ref{sec:conclusion}
contains our conclusions.

\newpage

\section{Interacting two-particle quantum walks and entanglement}
\label{sec:reviewmat}

As described in \cite{Omar2006, Berry2011}, 
a two-particle quantum walk takes place in the Hilbert space $\mathcal{H} =
\mathcal{H}_1 \otimes \mathcal{H}_2$, where $\mathcal{H}_{i} = (\mathcal{H}_v
\otimes \mathcal{H}_c)_{i}$ for particle $i$, with $v$ and $c$ representing the
position and coin space respectively.
Denoting the two-particle basis states as 
\begin{equation}
 \ket{v_i,v_k;c_j,c_l} = \ket{v_i,c_j}_1 \otimes \ket{v_k,c_l}_2 ,
\end{equation}
we can write the 
wavefunction describing the two-particle system as a linear combination of basis states
\begin{equation}
\label{eqn:pstate}
\ket{\psi} = \displaystyle\sum_{ik}\sum_{jl} a_{v_i,v_k;c_j,c_l}
\ket{v_i,v_k;c_j,c_l}.
\end{equation}
The normalization condition is expressible as:
\begin{equation}
\label{eqn:norm}
\displaystyle\sum_{ik}\sum_{jl} | a_{v_i,v_k;c_j,c_l} |^{2} = 1.
\end{equation}
The time evolution operator for the discrete-time two-particle quantum walk is 
\begin{equation}
\label{eqn:tprogopt}
U := S \cdot (\mathbbm{1} \otimes C ),
\end{equation}
where the shifting operator $ S = S_1 \otimes S_2 $  acts on the basis state as
\begin{equation}
S \ket{v_i,v_k;c_j,c_l} = \ket{v_j,v_l;c_i,c_k},
\end{equation}
and the coin operator $ C_{ik} $ is a $ d_i d_k \times d_i d_k $ unitary matrix
given by $ C_{ik} = (C_i)_1 \otimes (C_k)_2 $. Here $ d_i$ is the degree of
vertex $ i $, $ (C_i)_1 $ is the $ d_i \times d_i $ coin matrix for the vertex
$ v_i $, and $ (C_k)_2 $ is the $ d_k \times d_k $ coin matrix for the vertex $ v_k $.
The matrix $ C_{ik} $ acts on the $ d_i d_k $-dimensional coin space with the
basis states ordered in the following manner:
\begin{equation*}
\{
\ket{v_i,v_k;c_1,c_1},\ldots,\ket{v_i,v_k;c_1,c_{d_k}},\ket{v_i,v_k;c_2,c_1},\ldots,\ket{v_i,v_k;c_{d_i},c_{d_k}}
\}.
\end{equation*}

In this work, several initial states are considered. One initial state of the system is chosen to be an equal superposition of position
states on a graph with $ N $ vertices, 
\begin{equation}
\ket{\psi_0} = \displaystyle\sum_{i,k=1}^{n}\sum_{j=1}^{d_i}\sum_{l=1}^{d_k}
\frac{1}{N \sqrt{d_i d_k}} \ket{v_i,v_k;c_j,c_l}.
\end{equation}
Other initial states studied are random. Each coefficient $ a_{v_i,v_m;c_j,c_n} $ is chosen to be a randomly generated number from a uniform distribution over [0,1], and the state is then normalized using the normalization condition described in Eq.~(\ref{eqn:norm}).

The two-particle Grover coin operator is defined as $ C_{ik} = G(d_i) \otimes
G(d_k) $, where the one-particle Grover coin operator is given by
\begin{equation}
G_{\iota,\zeta}(d) = \frac{2}{d} - \delta_{\iota,\zeta}.
\end{equation}
where $\delta_{\iota,\zeta}$ is the Kronecker delta function. Such a two-particle coin operator is separable, and consequently the two particle
walkers evolve independently. To dynamically generate entanglement between the two particles,
we introduce a local interaction between the two particles, in the form of a
modification of the coin operator. A local interaction scheme perturbs the coin
operators only when both particles are simultaneously at the same node. We
define the effective coin operator as
\begin{equation}
\label{eqn:coin}
   C_{ik} = \left\{
     \begin{array}{lr}
       G(d_i) \otimes G(d_k) & : i \neq k \\
       C' & : i = k
     \end{array}
     \right.
\end{equation}
where $ C' $ is the perturbed coin operator. We consider the $ \phi $-Grover interaction scheme as introduced in \cite{Berry2011}, where $ C' = e^{i\phi} G(d_i) \otimes G(d_i) $.

The joint probability of particle 1 located at vertex $ v_i $ and particle 2
located at vertex $ v_k $ simultaneously, after $ t $ steps of the quantum walk
is
\begin{equation}
\label{eqn:prob}
P(i,k,t) = \displaystyle\sum_{jl} | \bra{v_i,v_k;c_j,c_l} (U)^t
\ket{\psi_0} |^{2}.
\end{equation}
The marginal probabilities for particle 1 and 2, assuming the particles are distinguishable, are obtained by summing the
joint probability over the position states of the other particle, i.e.
\begin{eqnarray}
\label{eqn:marginalprob}
P_1(i,t) = \displaystyle\sum_{k} P(i,k,t) & = & \displaystyle\sum_{k}
\displaystyle\sum_{jl} | \bra{v_i,v_k;c_j,c_l} (U)^t \ket{\psi_0} |^{2}, \notag \\
P_2(k,t) = \displaystyle\sum_{i} P(i,k,t) & = & \displaystyle\sum_{i}
\displaystyle\sum_{jl} | \bra{v_i,v_k;c_j,c_l} (U)^t \ket{\psi_0} |^{2}.
\end{eqnarray}

For a system in the pure state $ \ket{\psi} $ defined in Eq.~(\ref{eqn:pstate}), 
the reduced density matrix $ \rho_1 $ is obtained by tracing
the density matrix $ \rho = \ket{\psi}\bra{\psi} $ over subsystem 2 (namely both
position and coin states of particle 2). Using orthonormality, we obtain
\begin{equation}
\label{eqn:rho1def}
\rho_1 = \displaystyle \sum_{ik} \sum_{jl} b_{v_i,v_k;c_j,c_l}
\ket{v_i,c_j} \bra{v_k,c_l},
\end{equation}
where $ b_{v_i,v_k;c_j,c_l} = \sum_{m} \sum_{n} a_{v_i,v_m;c_j,c_n}
a_{v_k,v_m;c_l,c_n}^{*} $.  For such bipartite systems, entanglement can be
measured by the von Neumann entropy using the reduced density matrix of either
subsystem \cite{Mintert2005}
\begin{equation}
E(\ket{\psi}) = S(\rho_1) = -\mbox{Tr} \left( \rho_1 \mbox{ log}_2 ( \rho_1 )
\right),
\end{equation}
where $ \textbf{0} \mbox{ ln} ( \textbf{0} ) \equiv 0 $. Since the reduced density
matrix $ \rho_1 $ has real non-negative eigenvalues $ \lambda_i $ and the trace
is invariant under the similarity transformation, we can compute the
entanglement as
\begin{equation}
\label{eqn:entg}
E(\ket{\psi}) = - \displaystyle \sum_{i} \lambda_i \mbox{ log}_2 ( \lambda_i).
\end{equation}

\section{Entanglement time series and spectral analysis}
\label{sec:main}

Using the interacting coin operator defined by Eq.~(\ref{eqn:coin}), the entanglement between the two
walking particles varies in time and is also dependent on the interaction
strength $ \phi $. The corresponding power spectra of the entanglement time series are obtained via a discrete Fourier transform of a modified
time series.  We denote the (normalized) modulus square of the Fourier
transform as $ | \mathcal{E} |^2 $. The linear trend is subtracted from the
time series, and then multiplied by a tapered cosine window, defined by
\begin{equation}
	\label{eqn:window}
	\displaystyle
	w(n) = \left\{
	\begin{array}{lr}
		\frac{1}{2} \left( 1 + \mbox{cos}\left( \pi \left( \frac{2n}{\alpha(N-1)} - 1
		\right) \right) \right) & : 0 \leq n \leq \frac{\alpha(N-1)}{2} \\ 
		1 & : \frac{\alpha(N-1)}{2} \leq n \leq (N-1)(1-\frac{\alpha}{2}) \\ 
		\frac{1}{2} \left( 1 + \mbox{cos}\left( \pi \left( \frac{2n}{\alpha(N-1)} -
		\frac{2}{\alpha} + 1 \right) \right) \right) & : (N-1)(1-\frac{\alpha}{2})
		\leq n \leq N - 1
  	\end{array}
  	\right.
\end{equation}
where $ N $ is the width of the time series, $n = 0,1,\ldots,N-1 $, $ \alpha
N/2 $ is the width of the cosine lobes, and $ (1 - \alpha/2)N $ is the width of
the rectangle window. The window function is used to eliminate the
discontinuity at the endpoints of the time series, since the length of the
time series is not an exact multiple of the fundamental period, if it exists.
We then take the discrete Fourier transform of the modified time series.
The dependence of this power series on $ \phi $ can be visualised as a Feigenbaum bifurcation diagram. We plot the most prominent frequencies in the power series as a function of $ \phi $. Only frequencies with an amplitude in the top $5\%$ (black dots) and second $5\%$ (grey dots) of all amplitudes in the spectrum are shown. Frequencies with smaller amplitudes do not appear on this diagram. In this work, the parameter $ \alpha = 0.4 $ is used, which provides an effective reduction of noise in the frequency spectra.  A different value of $\alpha$ leads to a slightly different Feigenbaum diagram, but in any case it does not affect the overall appearance.  

%--------------------------------------------------

\subsection{Degree-regular graphs}

Degree-regular graphs have the same number of edges joined to each vertex, and as such tend to be very symmetric. First, we consider the highly symmetric and regular complete graphs on $n$ vertices, namely the $K_n$ graphs, with and without self-loops. Due to the symmetry and regularity of these graphs, when the system starts at an equal superposition of all states,  the joint probability distribution
defined in Eq.~(\ref{eqn:prob}) and the marginal probability distributions defined in Eq.~(\ref{eqn:marginalprob}) remain uniform at all times, which is independent of the $\phi$ value. That is, for a $K_n$ graph, $P(i,k,t) \equiv 1/{n^2} $ and $ P_1(i,t) \equiv P_2(k,t) \equiv P(j,t) \equiv 1/n $ for all $i$, $k$, $j$ and $t$.  This is a reflection of the symmetry of the $ K_n $ graphs, which are strongly regular, as well as vertex-, edge- and distance-transitive.
However, the entanglement between the two particles as defined in Eq.~(\ref{eqn:entg}) changes with time even for an initially equal superposition of all states, if they interact with a nonzero $\phi$ value. 

\begin{figure}[htp]
	\begin{center}
		\includegraphics[scale=0.7]{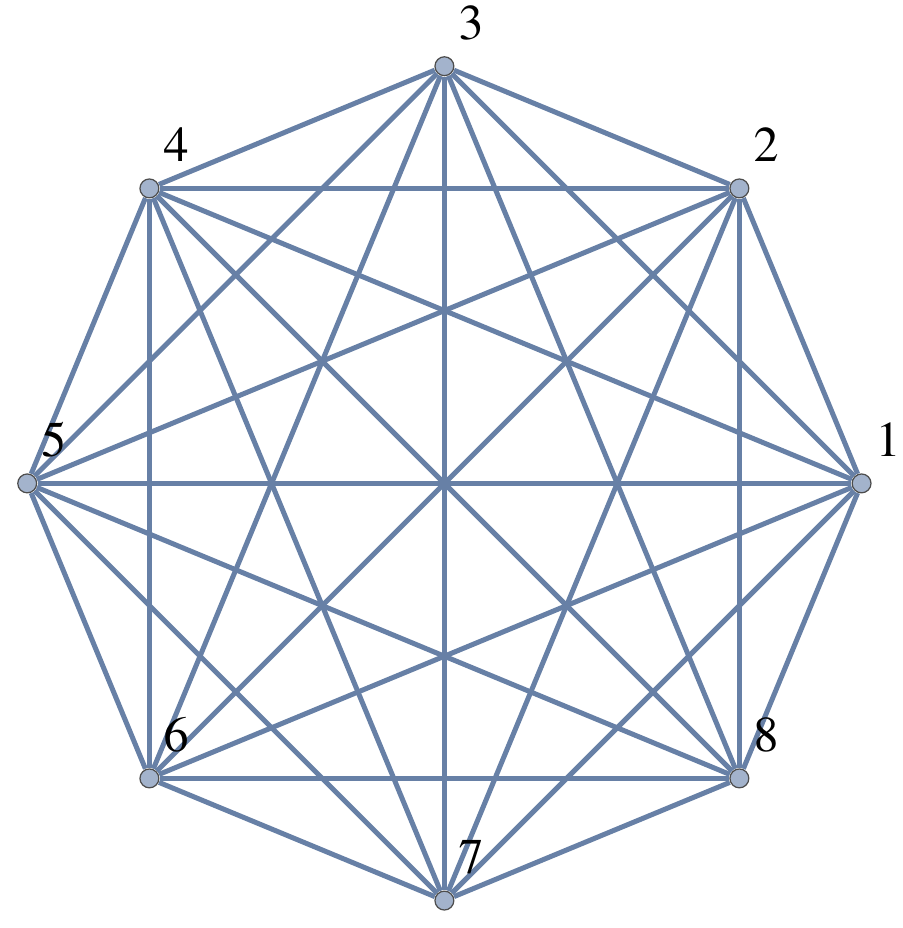}
	\end{center}
	\caption{$ K_8 $ graph without self loops.}
	\label{fig:K8graph}
\end{figure}

\begin{figure}[htp]
	\begin{center}
		\subfigure[t][$\phi=0.0$]{\label{fig:K8prob0}
		\includegraphics[scale=0.7]{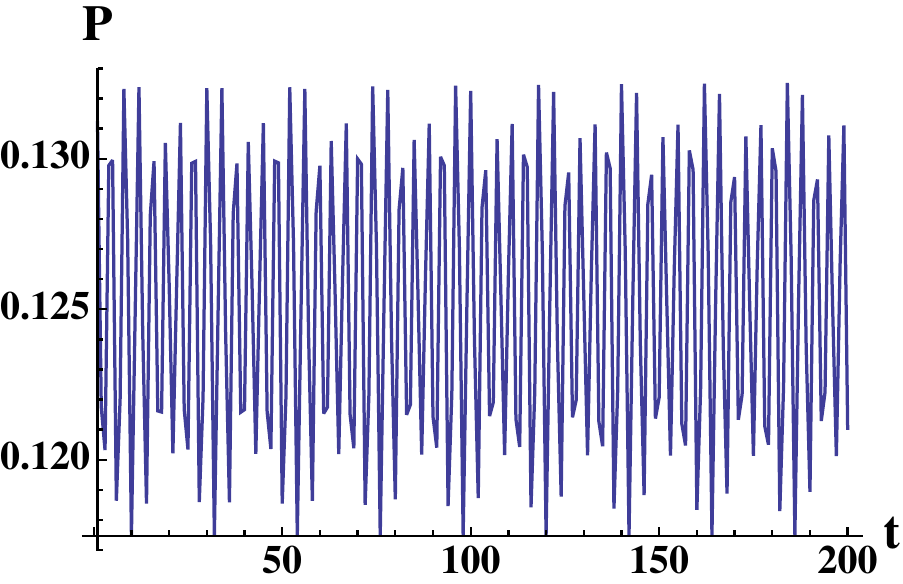}}
		\subfigure[t][$\phi=0.01\pi$]{\label{fig:K8prob0.01}
		\includegraphics[scale=0.7]{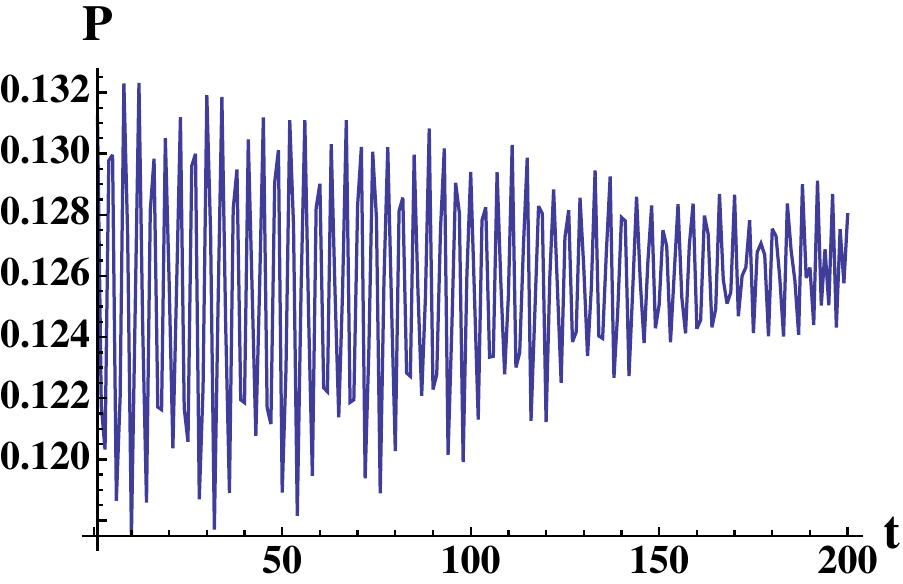}}\\
		\subfigure[t][$\phi=0.03\pi$]{\label{fig:K8prob0.03}
		\includegraphics[scale=0.7]{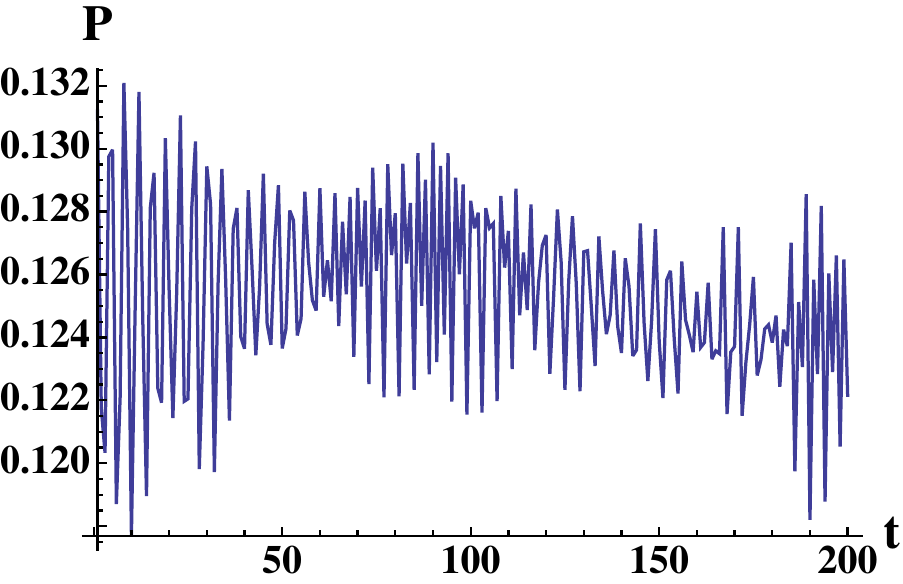}}
		\subfigure[t][$\phi=0.1\pi$]{\label{fig:K8prob0.1}
		\includegraphics[scale=0.7]{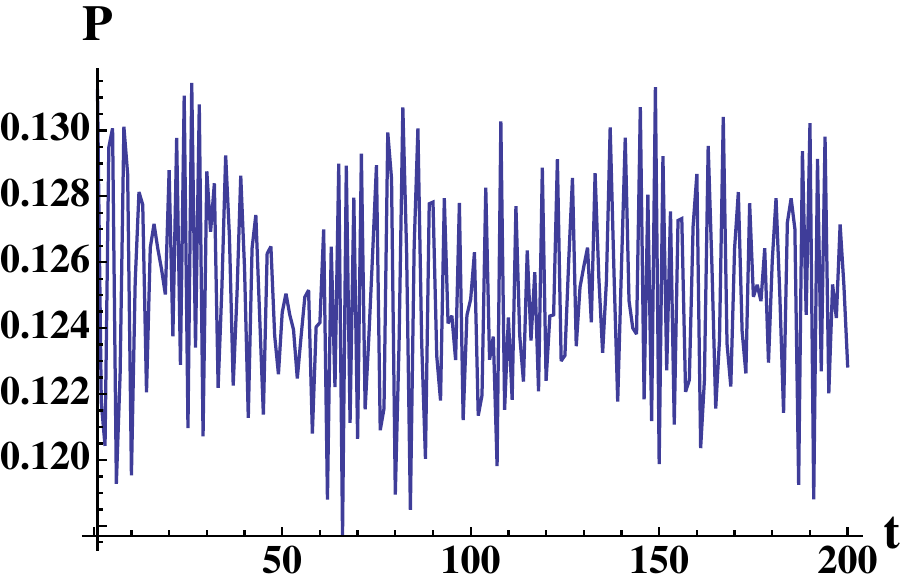}}
  	\end{center}
  	\caption{Marginal probability of a particle at vertex 1 for a
  	two-particle quantum walk on the $ K_8 $ graph without self-loops under  $ \phi
  	$-Grover interaction, with the initial state being random.}
  	\label{fig:K8prob}
\end{figure}

Since the probability and entanglement dynamics are very similar for these complete graphs, in the following we will only present the results for the $K_8$ graph without self loops (Figure \ref{fig:K8graph}).  We first show the marginal probability for a random initial state that breaks the symmetry. In this case the probability distributions vary in time and are different for different $\phi$ values, becoming more complex for larger $\phi$, as shown in Figure \ref{fig:K8prob}.

%\subsubsection{time series of probability distribution}

\begin{figure}[htp]
    \newpage
	\begin{center}
		\subfigure[t][$\phi=0.0$ (blue solid line), $\phi=0.02\pi$ (red dot-dashed line), $\phi=0.1\pi$ (green dotted line)]		{\label{fig:K8entsmall}
		\includegraphics[scale=0.8]{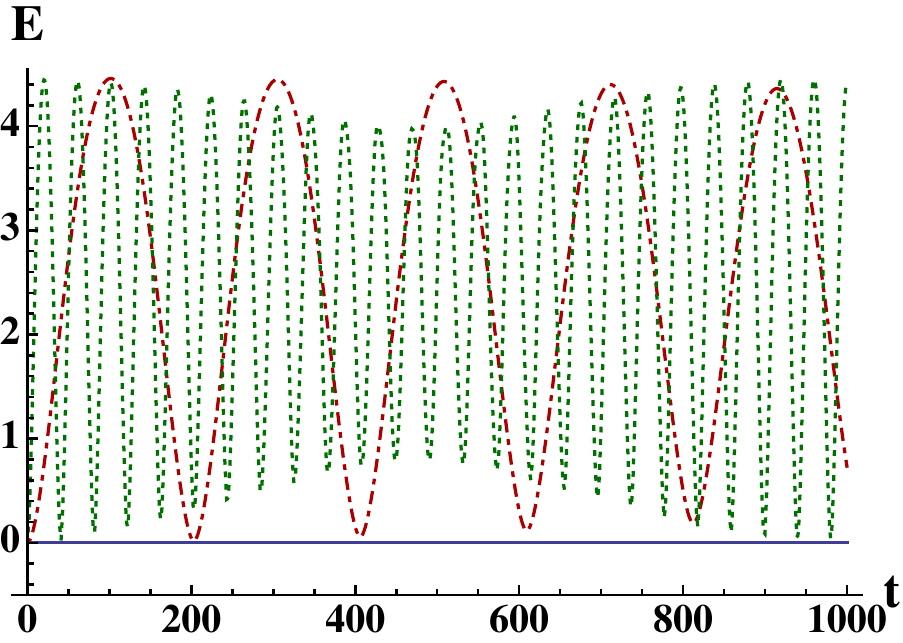}}
		\subfigure[t][$\phi=0.3\pi$]{\label{fig:K8ent0.3}
		\includegraphics[scale=0.8]{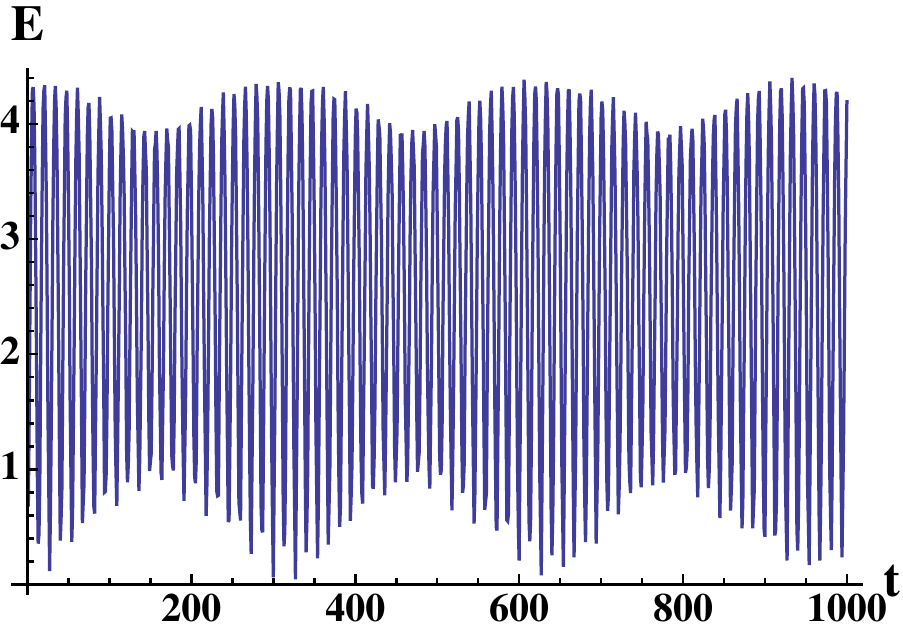}}\\
		\subfigure[t][$\phi=0.6\pi$]{\label{fig:K8ent0.6}
		\includegraphics[scale=0.8]{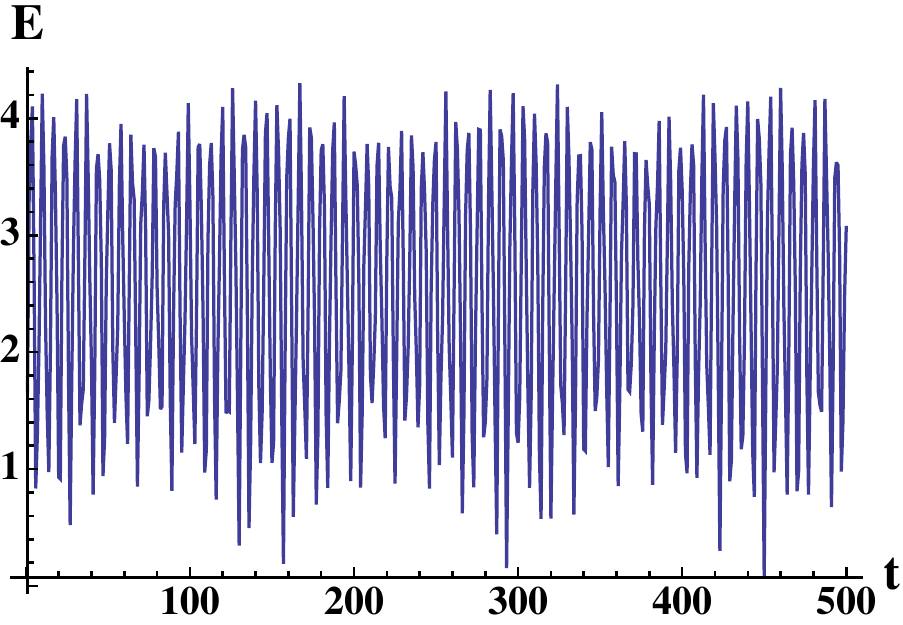}}
		\subfigure[t][$\phi=0.99\pi$]{\label{fig:K8ent0.99}
		\includegraphics[scale=0.8]{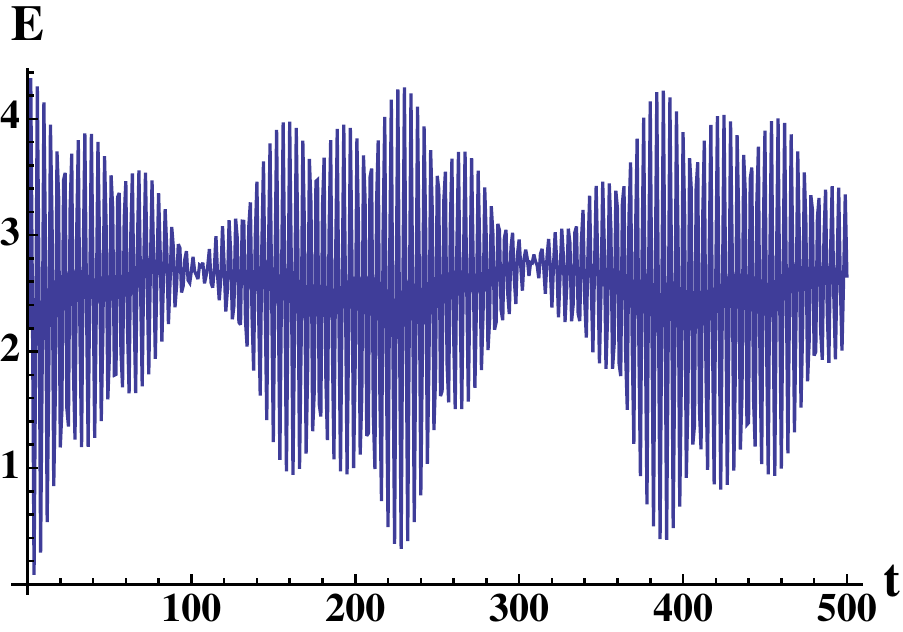}}
	\end{center}
  	\caption{Entanglement time series for a
  	two-particle quantum walk on the $ K_8 $ graph without self-loops under  $ \phi
  	$-Grover interaction, with an equal superposition initial state.}
  	\label{fig:K8ent}
\end{figure}

\begin{figure}[htp]
    \newpage
	\begin{center}
	    \subfigure[$K_8$]{\label{fig:K8Feig}
		\includegraphics[scale=0.7]{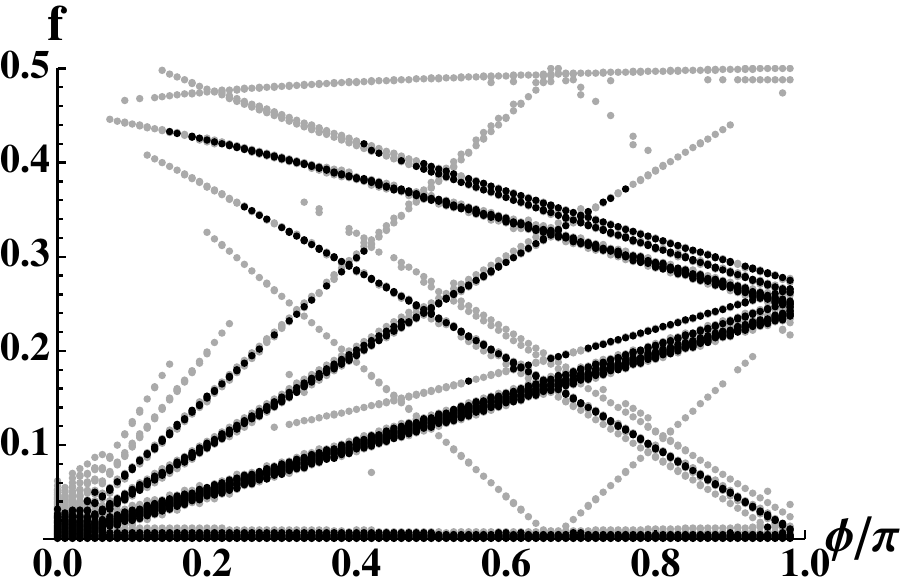}} \\
	    \subfigure[$Q_3$]{\label{fig:Q3Feig}
		\includegraphics[scale=0.7]{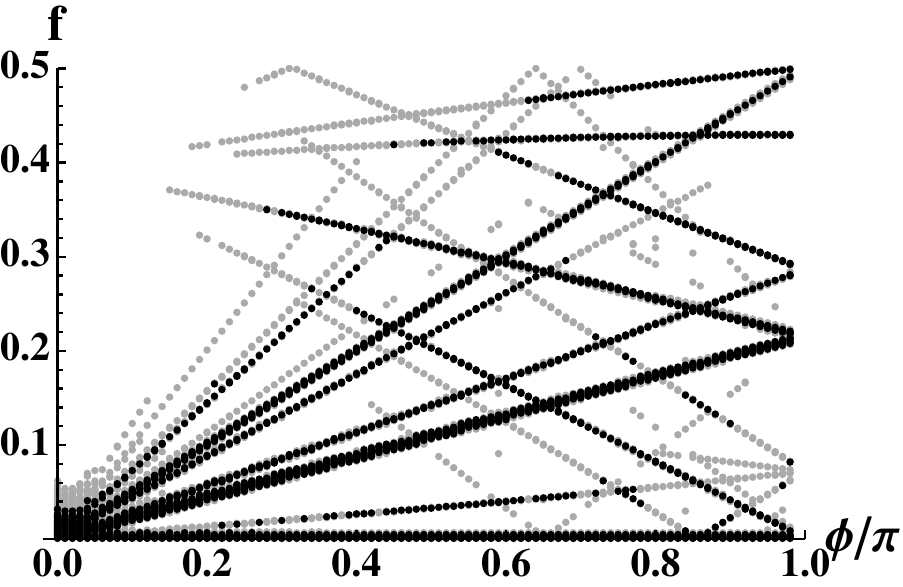}} \\
	    \subfigure[Joined 3CT2]{\label{fig:3CT2JFeig}
		\includegraphics[scale=0.7]{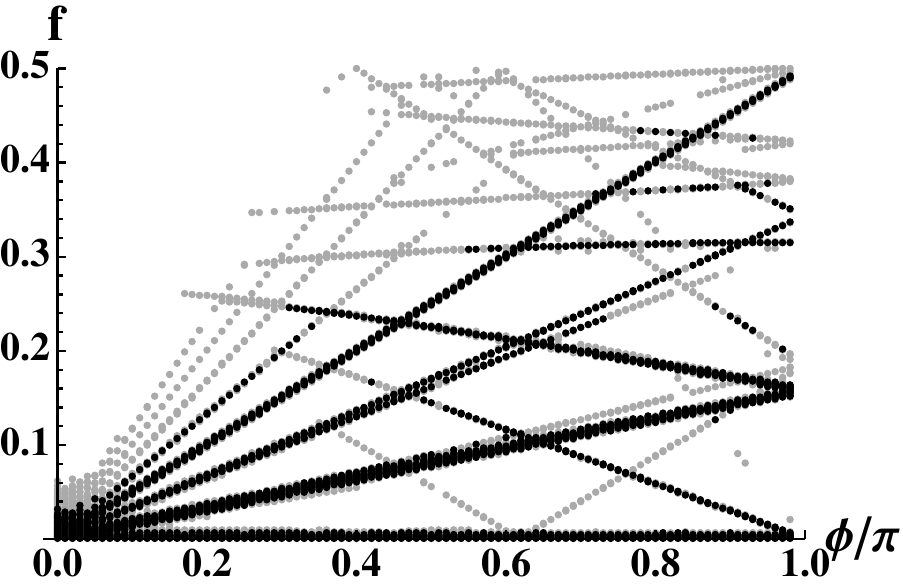}} \\
	\end{center}
	\caption{Feigenbaum diagrams of frequencies present in the entanglement
	time series of several regular graphs under $\phi$-Grover interaction, with an equal superposition initial state.}
	\label{fig:regFeig}
\end{figure}

We observe that the entanglement dynamics for the $ K_8 $ graph under the $\phi$-Grover interaction scheme are changing with time, even when the system begins at an equal superposition of all states, as shown in Figure \ref{fig:K8ent}.   Interesting to note is that for $\phi \rightarrow \pi$ the time series appears to form wave packets.  Nonetheless, the entanglement time series remain relatively simple and are oscillatory with the overall pattern repeating as $t \rightarrow \infty$.   The dynamics can be described by a number of frequencies, which increase for increasing $\phi$, as best seen in the Feigenbaum diagram shown in Figure \ref{fig:K8Feig}.  

%and a few more frequencies appear for larger $\phi$ values. 
%Only a few frequencies exist,
%, which is also true for the other graphs studied

\begin{figure}[htp]
	\begin{center}
		\includegraphics[scale=0.8]{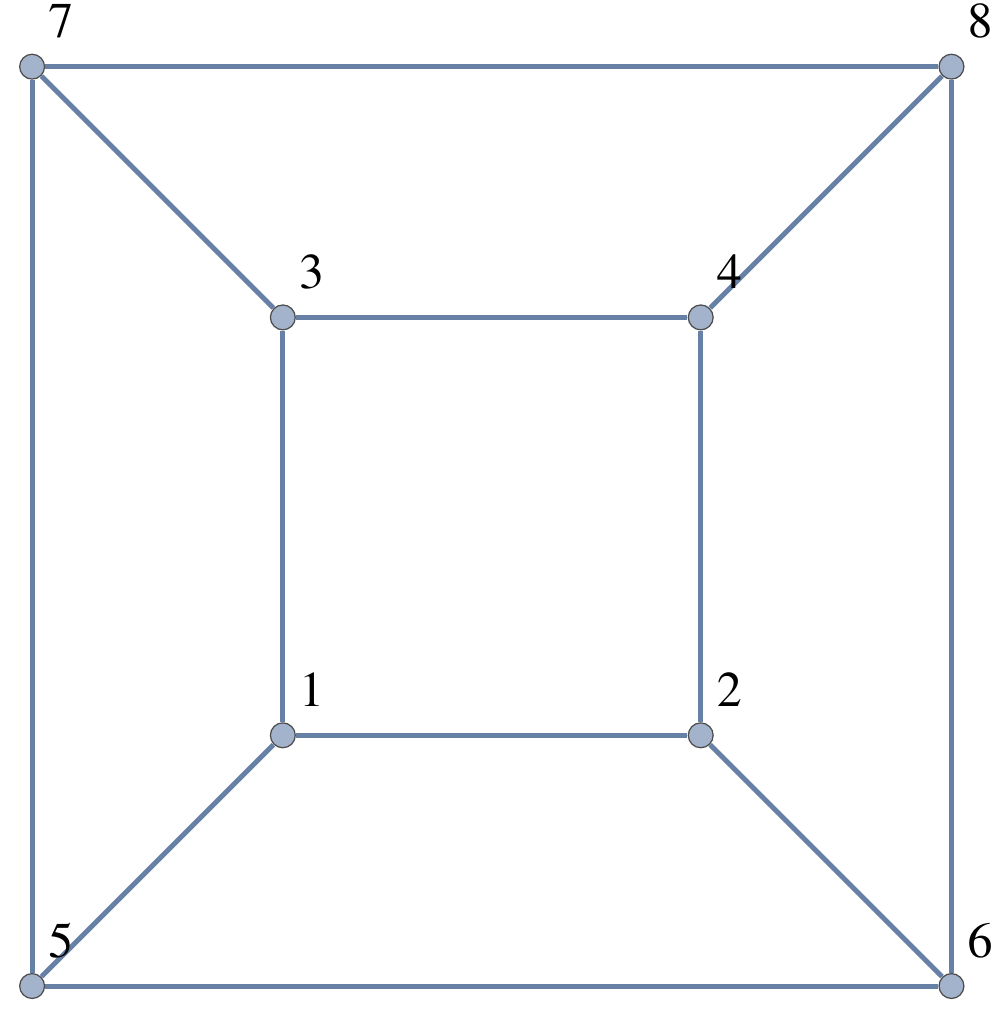}
	\end{center}
	\caption{3-dimensional hypercube graphs $Q_3$.}
	\label{fig:Q3graph}
\end{figure}

\begin{figure}[htp]
    \newpage
	\begin{center}
		\subfigure[t][$Q_3$ graph, $\phi=0.1\pi$ (blue solid line), $\phi=0.3\pi$ (red dot-dashed line)]{\label{fig:Q3entsmall}
		\includegraphics[scale=0.8]{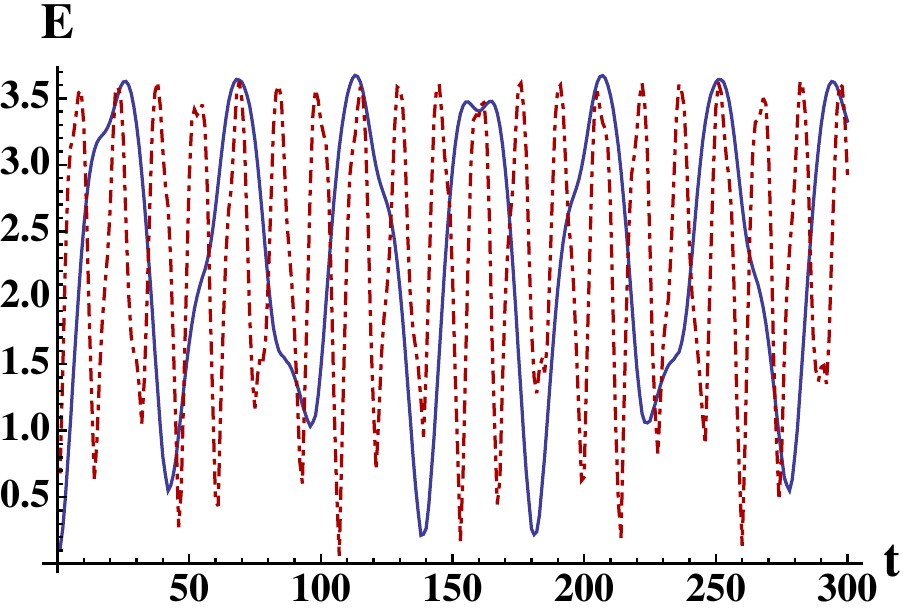}}
		\subfigure[t][$Q_3$ graph, $\phi=0.6\pi$ (blue solid line), $\phi=0.99\pi$ (red dot-dashed line)]{\label{fig:Q3entlarge}
		\includegraphics[scale=0.8]{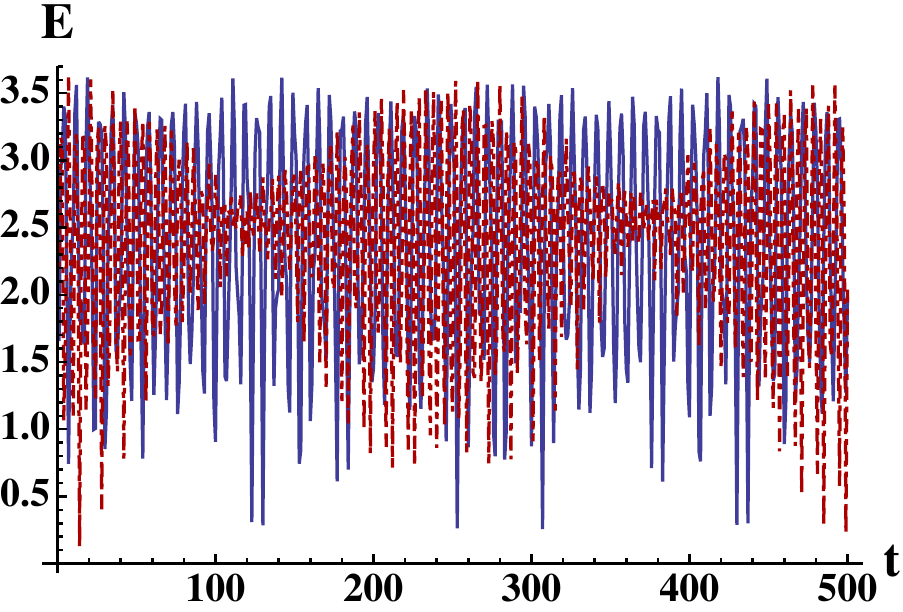}}
	    \subfigure[t][Joined 3CT2 graph, $\phi=0.1\pi$ (blue solid line), $\phi=0.3\pi$ (red dot-dashed line)]{\label{fig:3CT2Jentsmall}
		\includegraphics[scale=0.8]{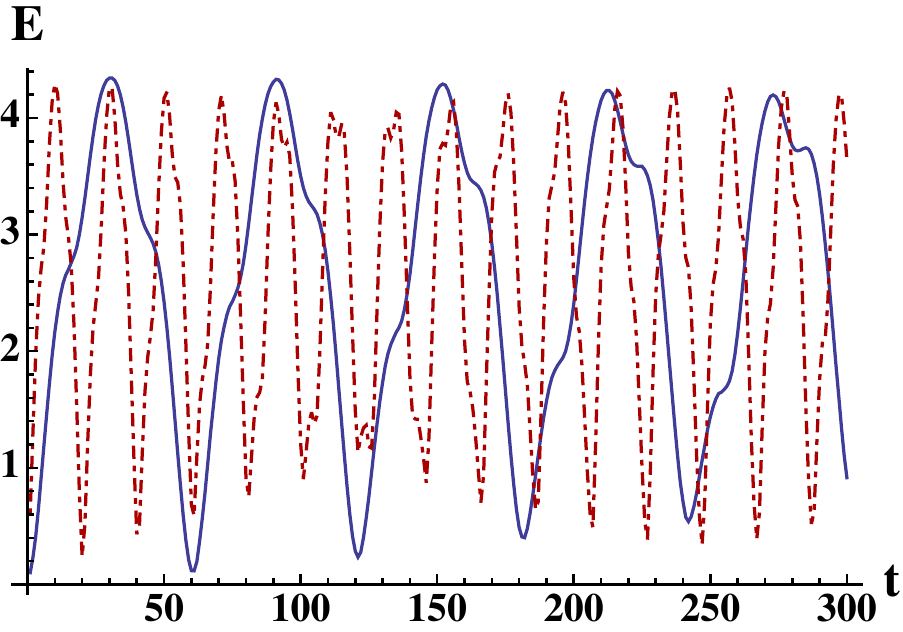}}
		\subfigure[t][Joined 3CT2 graph, $\phi=0.6\pi$ (blue solid line), $\phi=0.99\pi$ (red dot-dashed line)]{\label{fig:3CT2Jentlarge}
		\includegraphics[scale=0.8]{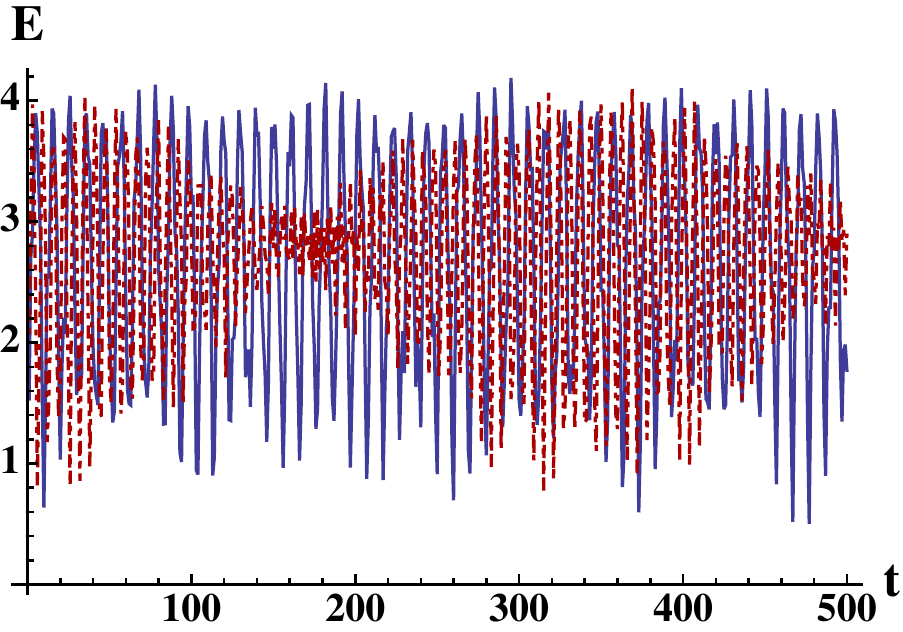}}
	\end{center}
  	\caption{Entanglement time series for a
  	two-particle quantum walk on the $Q_3$ and joined 3CT2 graphs under  $ \phi
  	$-Grover interaction, with an equal superposition initial state. }
  	\label{fig:Q3_3CT2J_ent}
\end{figure}

Another family of very symmetric regular graphs is the hypercube graphs $Q_n$. Two-particle quantum walks on the 3-dimensional hypercube graph (Figure \ref{fig:Q3graph}) were considered. The $Q_3$ graph has the same dimension as the $K_8$ graph, so the two are comparable.
As can be seen in Figure \ref{fig:Q3_3CT2J_ent}, the entanglement dynamics of the $Q_3$ graph differ from that of the $K_8$ graph, but follow the same trend of increasing complexity as $\phi$ increases. The entanglement time series appears less regular for this graph than for the $K_8$ graph, reflecting a reduced degree of symmetry.

\begin{figure}[htp]
	\begin{center}
		\includegraphics[scale=0.8]{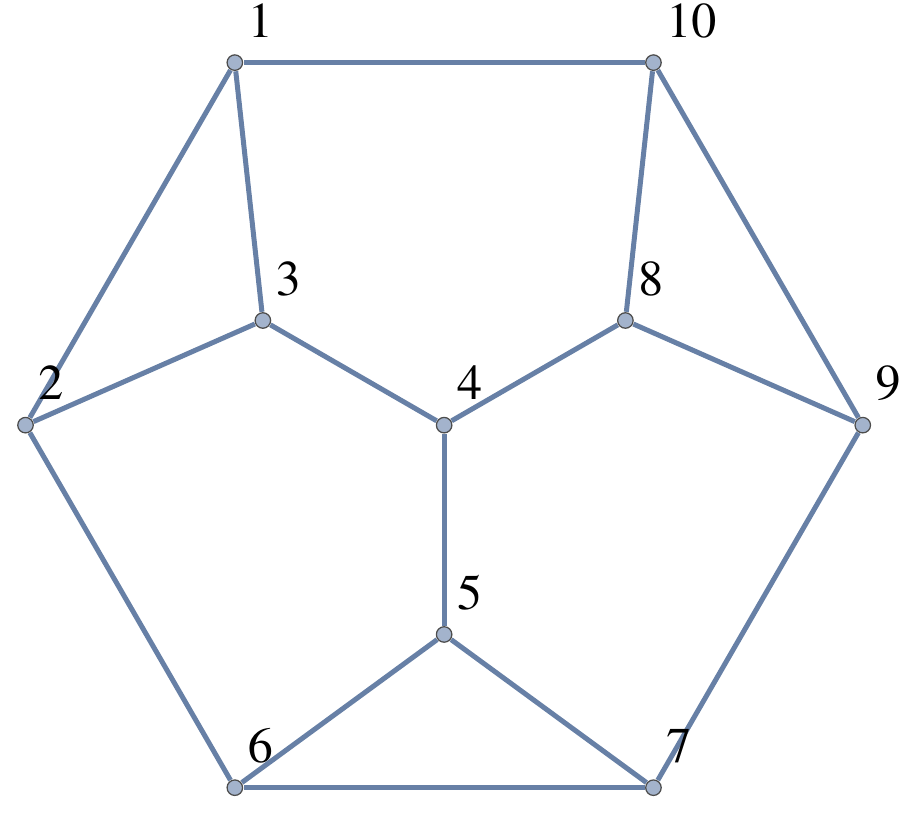}
	\end{center}
	\caption{Joined 2nd generation 3-Cayley Tree.}
	\label{fig:3CT2Jgraph}
\end{figure}

Walks on the joined 2nd generation 3-Cayley tree (3CT2) shown in Figure \ref{fig:3CT2Jgraph} were also studied. Again, the complexity of the time series increases with increasing $\phi$, and these time series again appear more complex than those for the $K_8$ graph (Figure \ref{fig:Q3_3CT2J_ent}).  Comparing the Feigenbaum diagrams for the entanglement time series of the $K_8$, $Q_3$ and joined 3CT2 graphs (Figure \ref{fig:regFeig}), it can be seen that for all of these graphs the frequency spectra behave similarly as $\phi$ changes. All have a few prominent frequencies for lower values of $\phi$, and these frequencies change on the whole linearly with $\phi$. The dynamics become much more complex with increasing $\phi$, with more frequencies emerging. The $Q_3$ and joined 3CT2 graphs have many more frequencies than the $K_8$ graph, especially for the higher values of $\phi$.

\begin{figure}[htp]
    \newpage
	\begin{center}
		\subfigure[$K_8$ graph]{\label{fig:K8perturb}
		\includegraphics[scale=0.7]{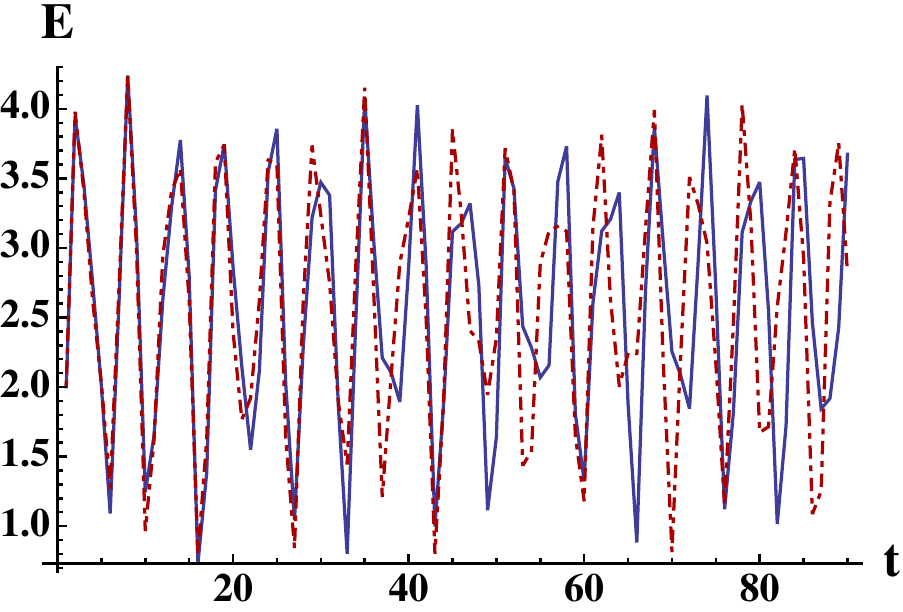}}
		\subfigure[$Q_3$ graph]{\label{fig:Q3perturb}
		\includegraphics[scale=0.7]{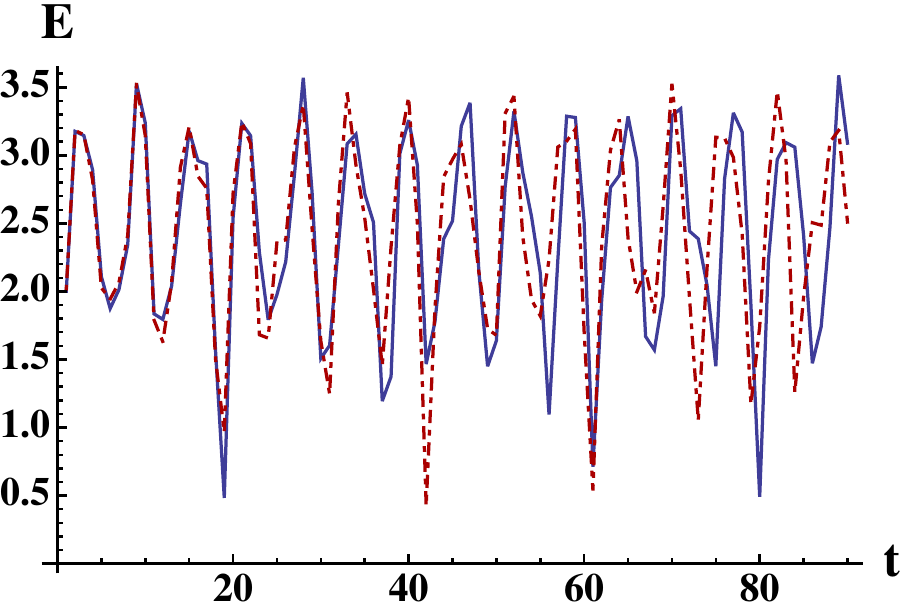}}
		\subfigure[Joined 3CT2 graph]{\label{fig:3CT2Jperturb}
		\includegraphics[scale=0.7]{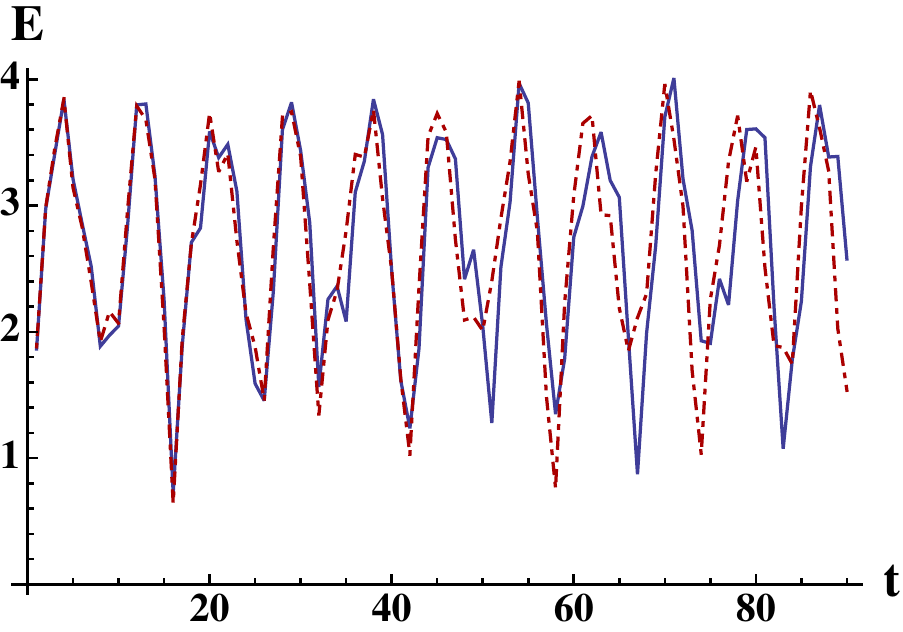}}
	\end{center}
	\caption{Time series of entanglement for $ \phi = 0.75\pi $ (blue, solid) and $\phi = 0.76\pi $ (green, dotted) for a two-particle quantum walk under $ \phi$-Grover interaction beginning in an equal superposition initial state.}
	\label{fig:reg_perturb}
\end{figure}

\begin{figure}[htp]
	\begin{center}
	    \subfigure[Equal superposition initial state]{\label{fig:Q3Feigequal}
		\includegraphics[scale=0.8]{Q3Feig-eps-converted-to.pdf}} 
	    \subfigure[Random initial state]{\label{fig:Q3Feigrand}
		\includegraphics[scale=0.8]{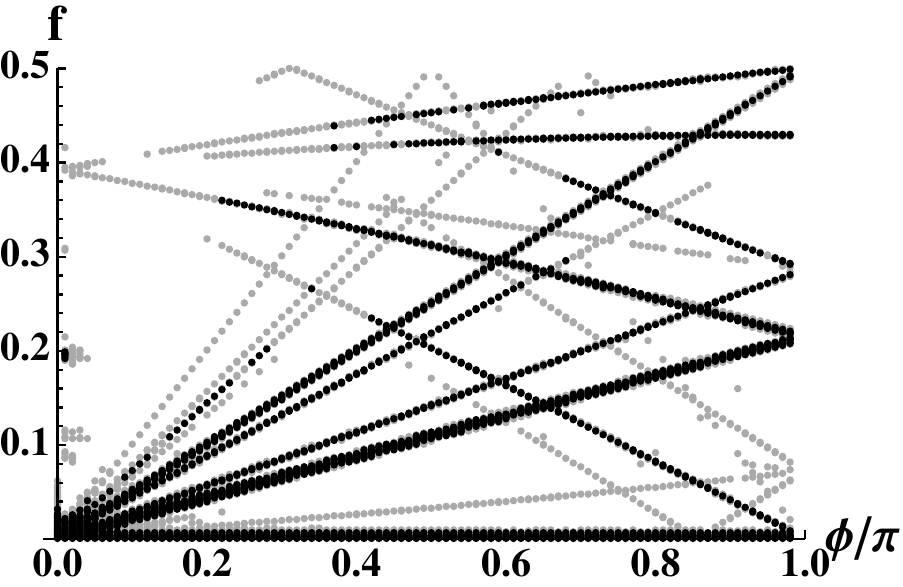}} \\
	\end{center}
	\caption{Feigenbaum diagrams of frequencies present in the entanglement
	time series of the $Q_3$ graph under $\phi$-Grover interaction.}
	\label{fig:comp_initial_Feig}
\end{figure}

As shown in Figure \ref{fig:reg_perturb}, the entanglement is sensitive to small perturbations in the $\phi$ parameter for every regular graph studied. This is true for every initial state considered. However, the marginal single-particle probability time series are not so sensitive to small changes in $\phi$. For the equal superposition initial state, the probability time series remains uniform, independent of $\phi$. For the random initial state, probability time series are complex but essentially stable. On the other hand, the entanglement time-series are not sensitive to the initial state of the system. The various initial states give similar entanglement time series, appearing to have the same frequencies and general shape. The Feigenbaum diagrams for each of the regular graphs discussed above are almost identical for both the equal superposition and a random initial state, as shown in Figure \ref{fig:comp_initial_Feig}.

\subsection{Non-degree-regular graphs}

\begin{figure}[htp]
      \newpage
	\begin{center}
	\subfigure[t][$K_8$ with 10/28 edges removed]{\label{fig:K8irrgraph}
		\includegraphics[scale=0.8]{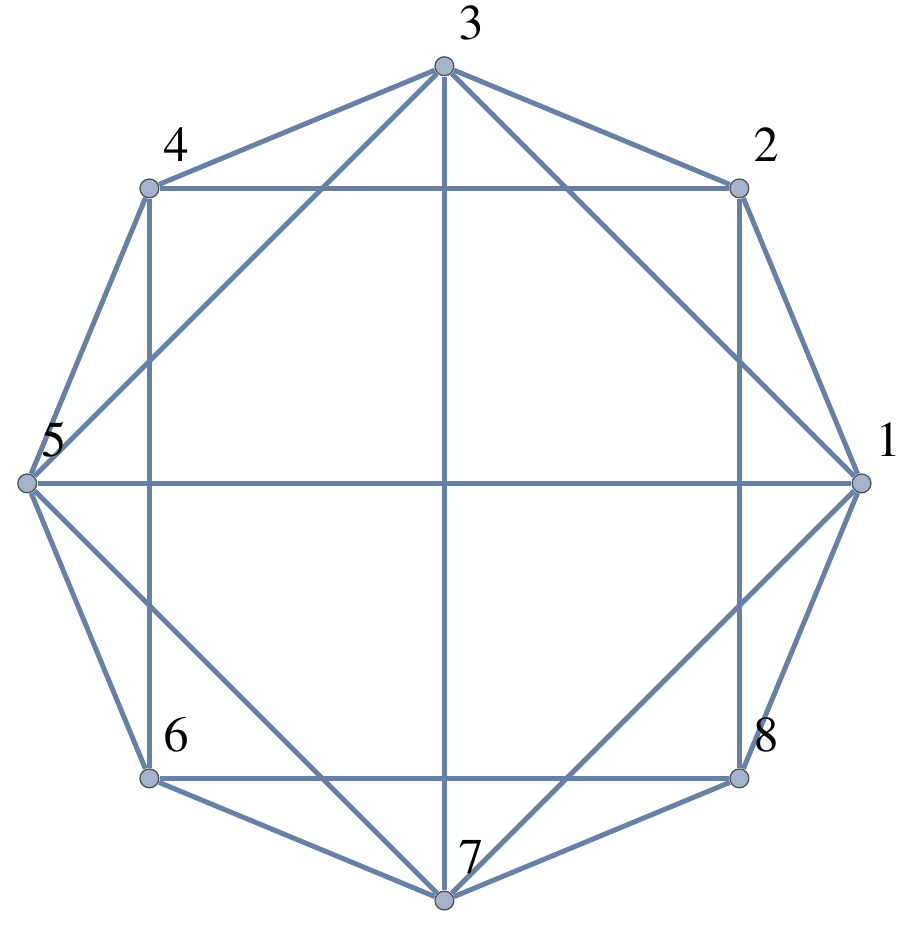}}
	\subfigure[t][$Q_3$ with 4/12 edges removed.]{\label{fig:Q3irrgraph}
		\includegraphics[scale=0.8]{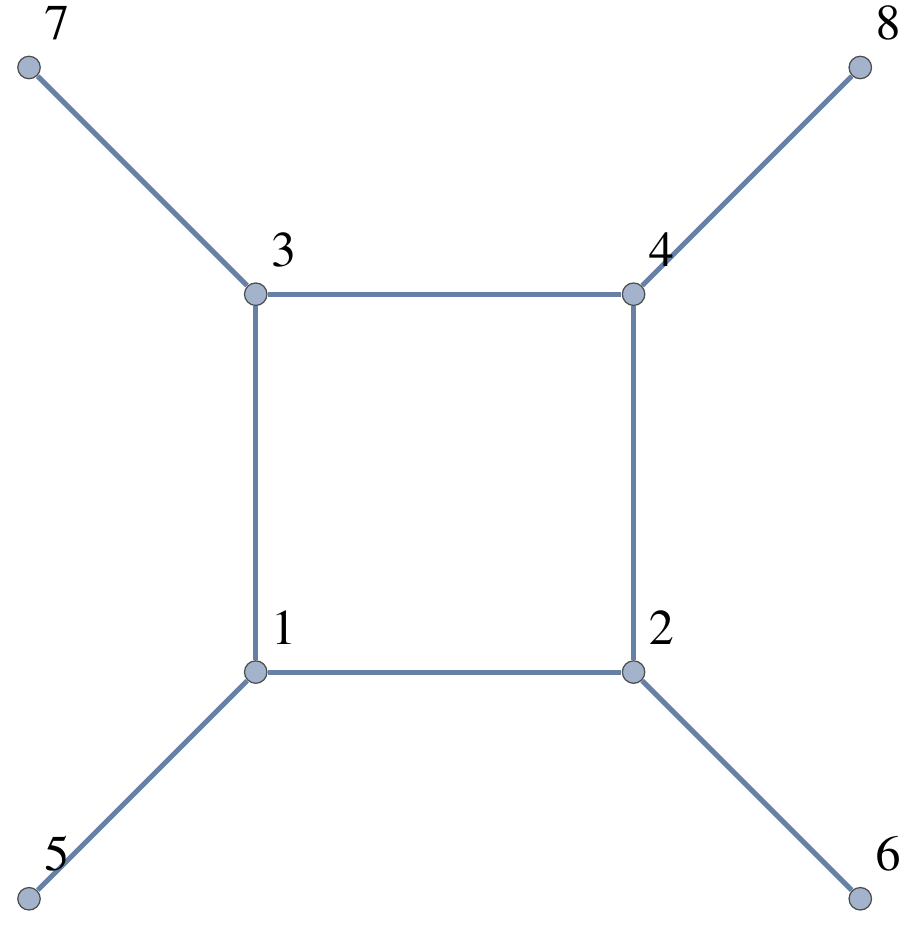}}
	\subfigure[t][ 2nd generation 3-Cayley tree]{\label{fig:3CT2graph}
		\includegraphics[scale=0.8]{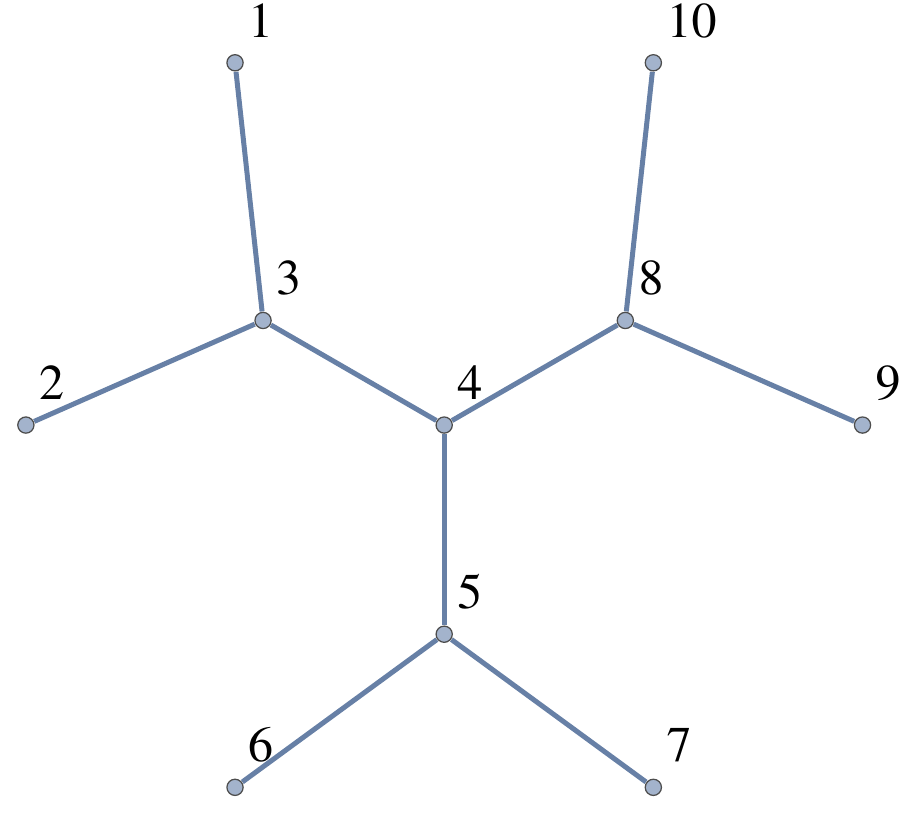}}
	\end{center}
	\caption{Non-degree-regular counterparts of regular graphs.}
	\label{fig:irreg_graphs}
\end{figure}

We now consider non-degree-regular graphs, and for comparison purposes these are all similar to the degree-regular graphs considered above, with several edges removed so that not all vertices are joined to the same number of edges. The irregularity of these graphs introduce more complex dynamics into the system, making them good candidates to study the complexity in the time-evolution of probability distributions and quantum correlations within the system with varying interaction strength $ \phi $.
The non-degree-regular graphs studied in this work are the unjoined 2nd generation 3-Cayley tree, a modified $K_8$ graph and a modified $Q_3$ graph, all with between $30\%$ and $40\%$ of their edges removed (Figure \ref{fig:irreg_graphs}). These graphs still have a high degree of symmetry, as their regular counterparts do.

\begin{figure}[htp]
    \newpage
	\begin{center}
		\subfigure[t][$\phi=0.0$]{\label{fig:K8irrprob0}
		\includegraphics[scale=0.8]{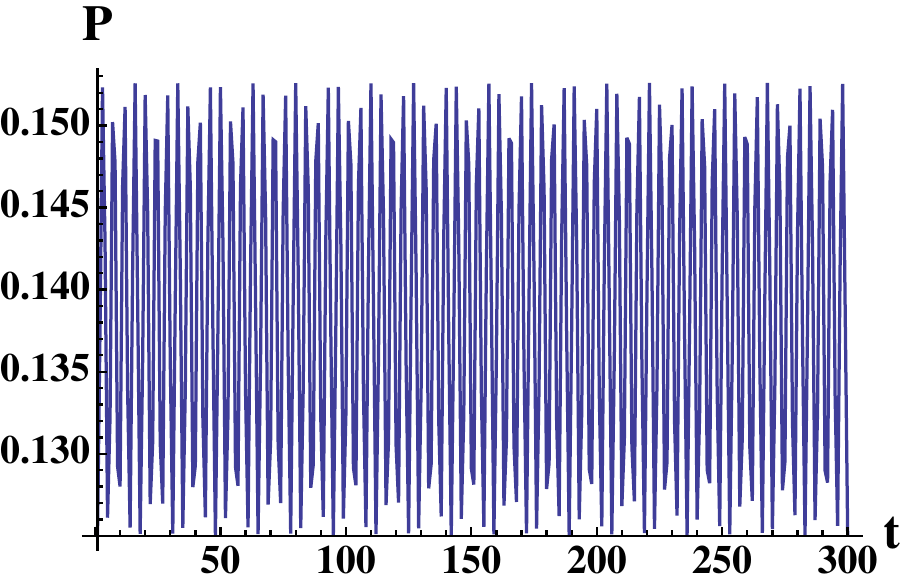}}
		\subfigure[t][$\phi=0.01\pi$]{\label{fig:K8irrprob0.01}
		\includegraphics[scale=0.8]{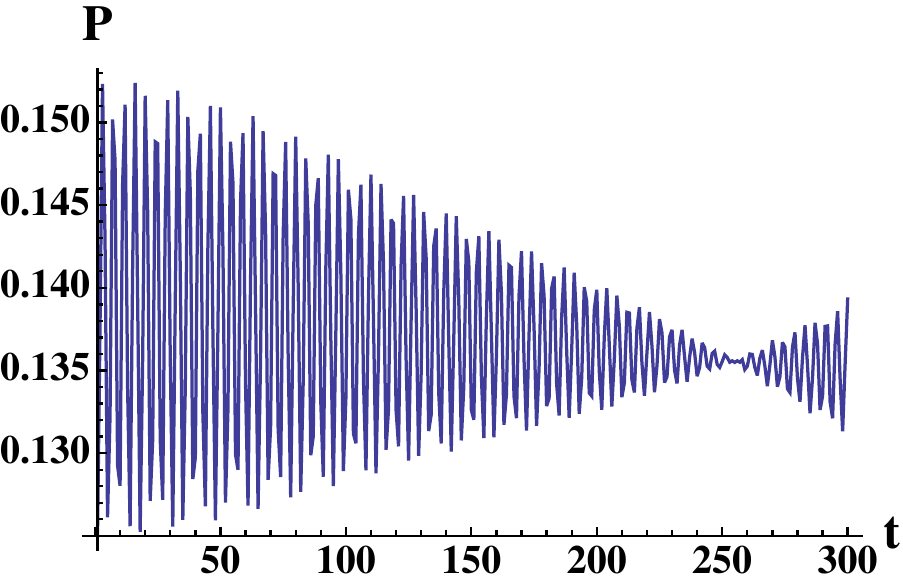}}\\
		\subfigure[t][$\phi=0.03\pi$]{\label{fig:K8irrprob0.03}
		\includegraphics[scale=0.8]{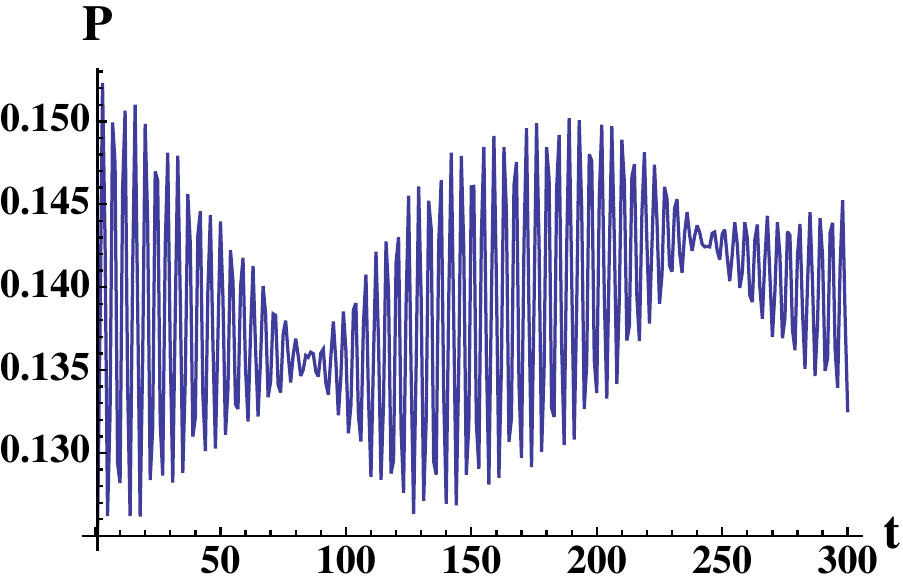}}
		\subfigure[t][$\phi=0.1\pi$]{\label{fig:K8irrprob0.1}
		\includegraphics[scale=0.8]{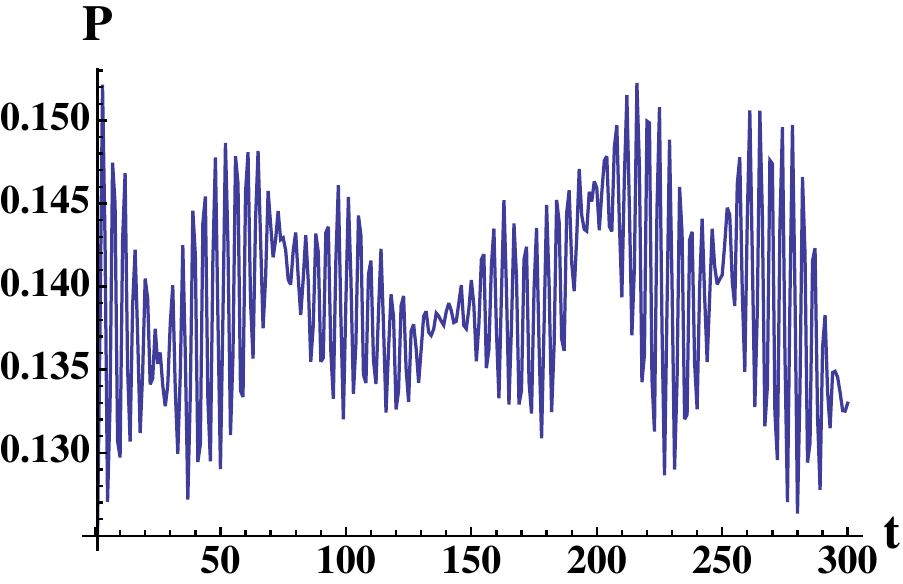}}
  	\end{center}
  	\caption{Marginal probability of a particle at vertex 1 for a
  	two-particle quantum walk on the modified $ K_8 $ graph without self-loops under  $ \phi
  	$-Grover interaction, with the initial state being an equal superposition of all states.}
  	\label{fig:K8irrprob}
\end{figure}

These graphs have different probability time series to their regular analogues. The single-particle probability is no longer uniform for the equal superposition equal state, but is periodic for very low values of $\phi$, and increasingly complex for larger $\phi$. Figure \ref{fig:K8irrprob} shows these time series for the modified $K_8$ graph. When the system starts in a random initial state, the time series are no longer very regular for lower values of $\phi$.

\begin{figure}[htp]
\newpage
	\begin{center}
		\subfigure[t][$\phi=0.0$ (blue solid line), $\phi=0.02\pi$ (red dot-dashed line), $\phi=0.1\pi$ (green dotted line)]{\label{fig:K8irrentsmall}
		\includegraphics[scale=0.8]{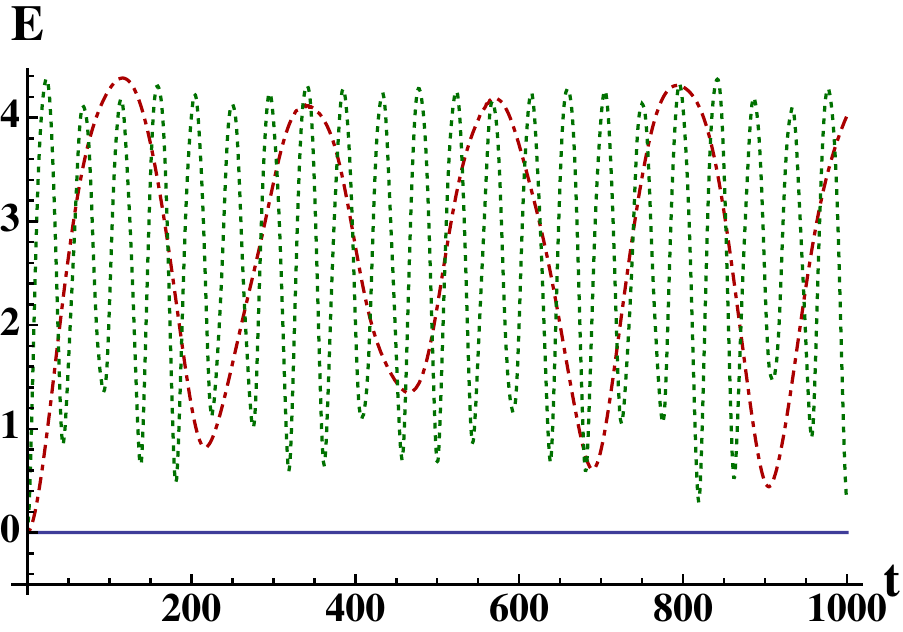}}
		\subfigure[t][$\phi=0.3\pi$]{\label{fig:K8irrent0.3}
		\includegraphics[scale=0.8]{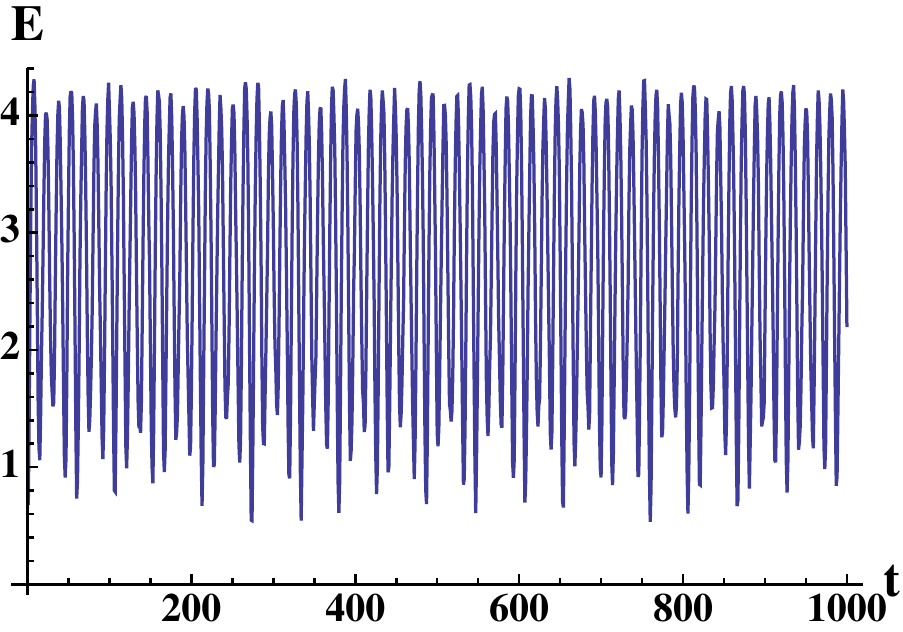}}\\
		\subfigure[t][$\phi=0.6\pi$]{\label{fig:K8irrent0.6}
		\includegraphics[scale=0.8]{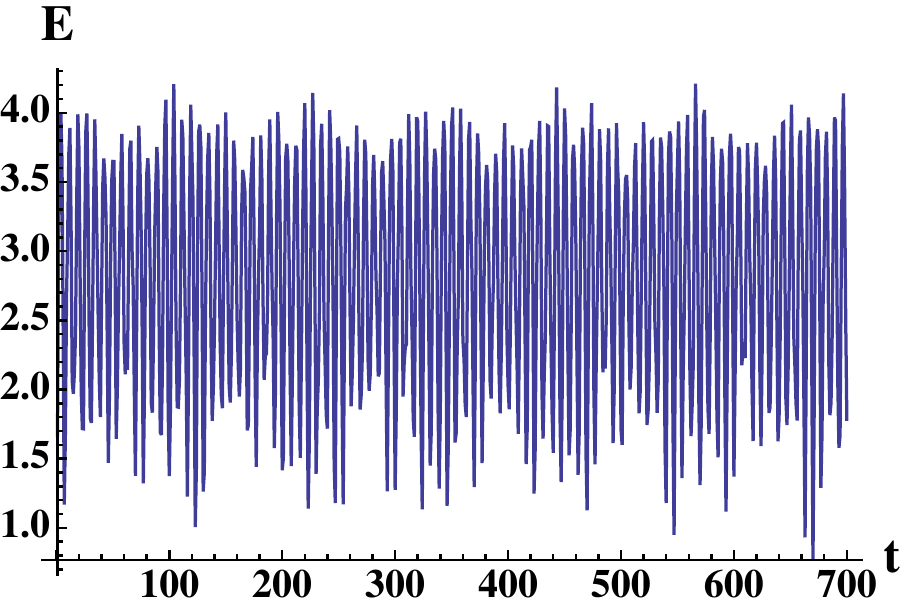}}
		\subfigure[t][$\phi=0.99\pi$]{\label{fig:K8irrent0.99}
		\includegraphics[scale=0.8]{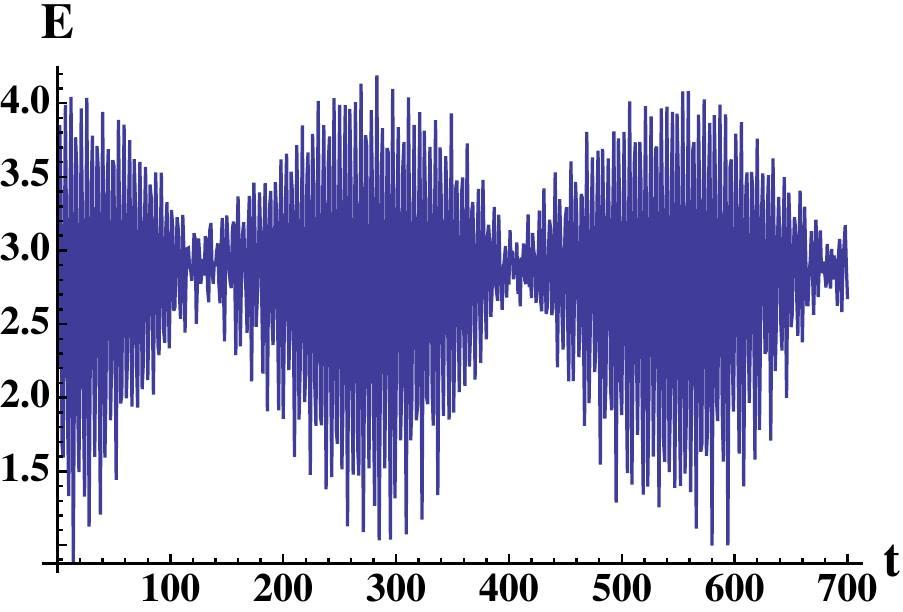}}
	\end{center}
  	\caption{Entanglement time series for a
  	two-particle quantum walk on the modified $ K_8 $ graph without self-loops under  $ \phi
  	$-Grover interaction, with an equal superposition initial state.}
  	\label{fig:K8irrent}
\end{figure}

\begin{figure}[htp]
    \newpage
	\begin{center}
		\subfigure[t][Modified $Q_3$ graph, $\phi=0.1\pi$ (blue solid line), $\phi=0.3\pi$ (red dot-dashed line)]{\label{fig:Q3irrentsmall}
		\includegraphics[scale=0.8]{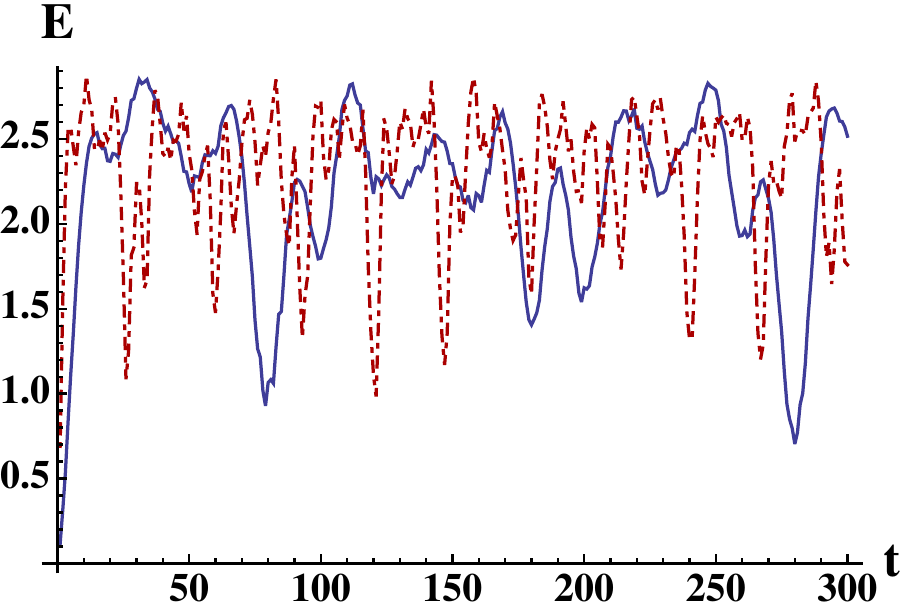}}
		\subfigure[t][Modified$Q_3$ graph, $\phi=0.6\pi$ (blue solid line), $\phi=0.99\pi$ (red dot-dashed line)]{\label{fig:Q3irrentlarge}
		\includegraphics[scale=0.8]{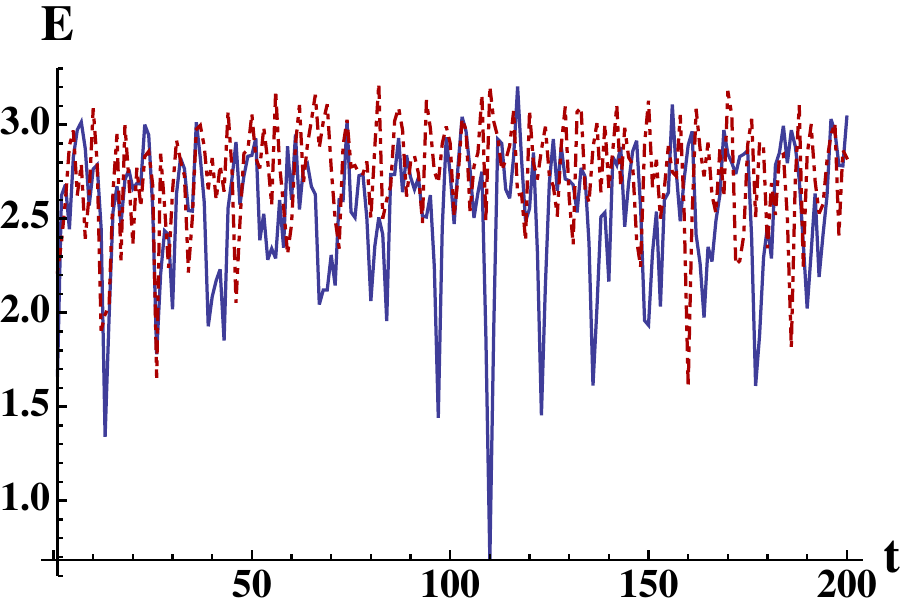}}
		\subfigure[t][Unjoined 3CT2 graph, $\phi=0.1\pi$ (blue solid line), $\phi=0.3\pi$ (red dot-dashed line)]{\label{fig:3CT2entsmall}
		\includegraphics[scale=0.8]{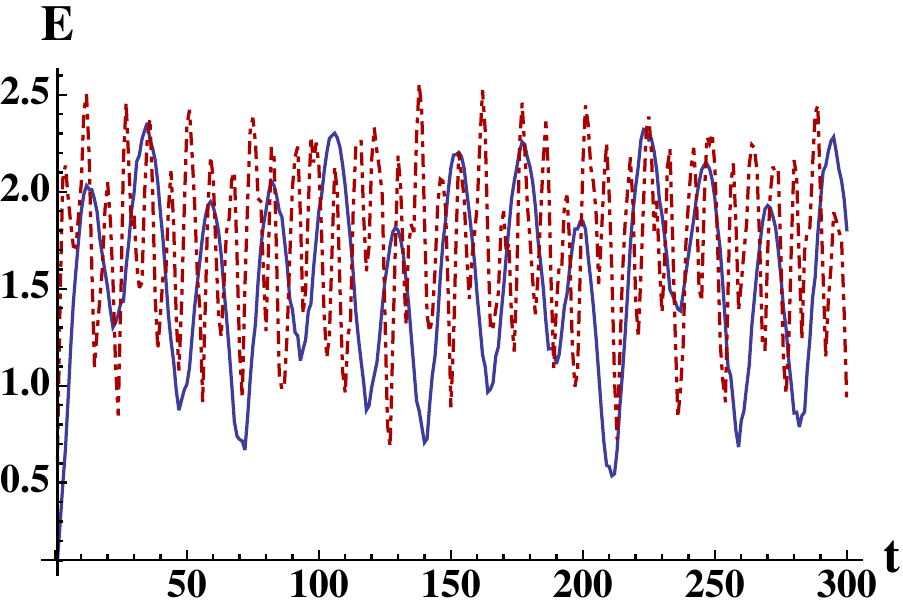}}
		\subfigure[t][Unjoined 3CT2 graph, $\phi=0.6\pi$ (blue solid line), $\phi=0.99\pi$ (red dot-dashed line)]{\label{fig:3CT2entlarge}
		\includegraphics[scale=0.8]{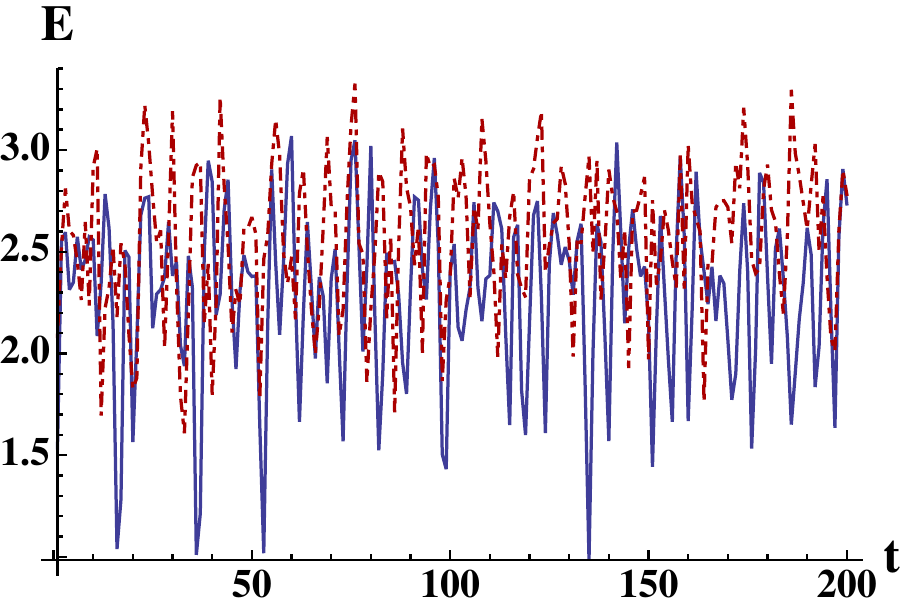}}
	\end{center}
  	\caption{Entanglement time series for a
  	two-particle quantum walk on the modified $Q_3$ and unjoined 3CT2 graphs under  $ \phi
  	$-Grover interaction, with an equal superposition initial state. }
  	\label{fig:Q3irr_3CT2_ent}
\end{figure}

We now consider the entanglement dynamics of two-particle quantum walks along these irregular graphs.
There is no entanglement between the two particles for $ \phi = 0$, which corresponds to having no interaction between the two particles. With weak interaction at small values of $ \phi $, the
entanglement time series is periodic with a fixed number of frequencies. As we increase $ \phi $, we observe a transition to irregular/quasi-periodic dynamics as more frequencies are
introduced further into the system. Figure \ref{fig:K8irrent} illustrates this for the modified $K_8$ graph, while some entanglement time series for the other two irregular graphs can be seen in Figure \ref{fig:Q3irr_3CT2_ent}.

\begin{figure}[htp]
	\begin{center}
		\subfigure[t][$K_8$]{\label{fig:K8regirregcomp}
		\includegraphics[scale=0.5]{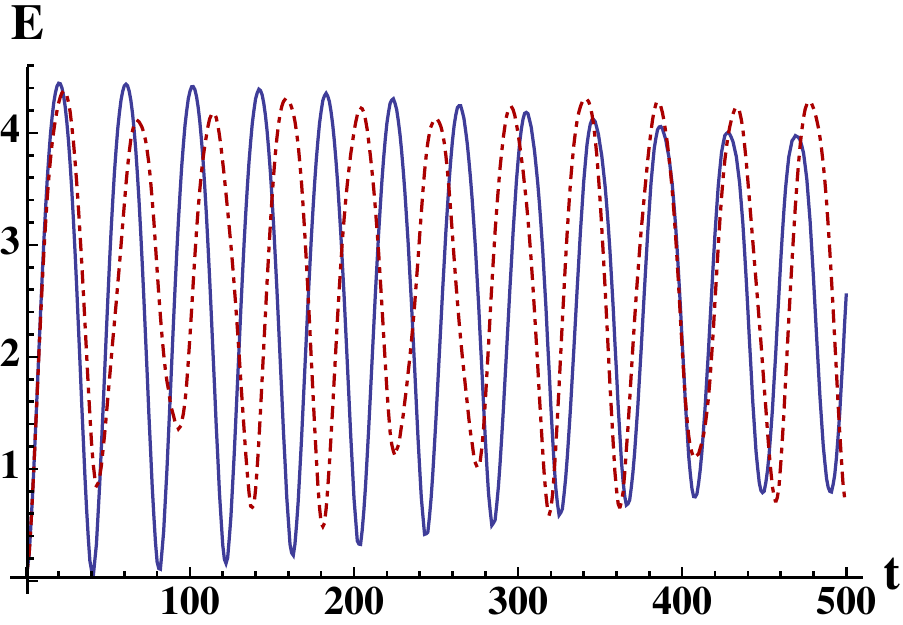}}
		\subfigure[t][$Q_3$]{\label{fig:Q3regirregcomp}
		\includegraphics[scale=0.5]{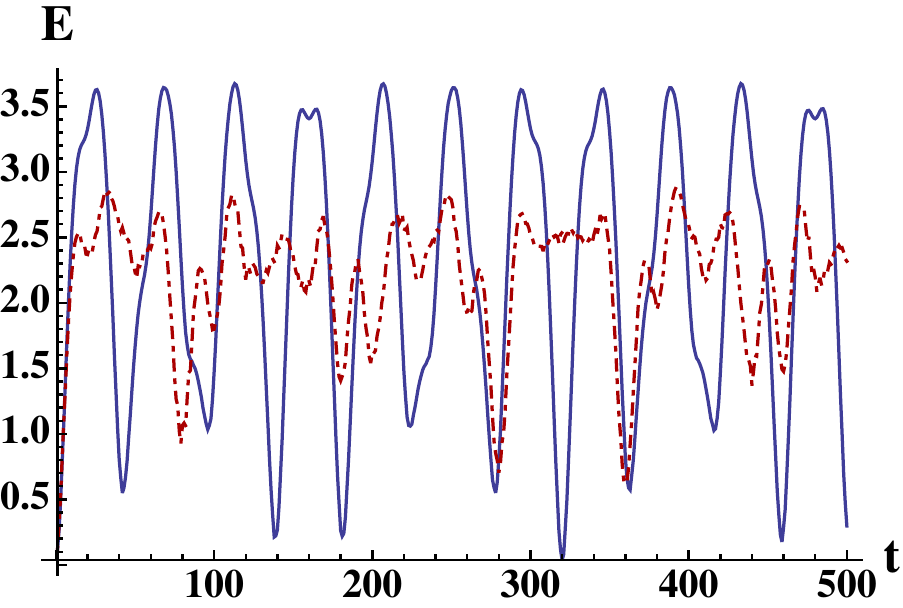}}
		\subfigure[t][Cayley tree]{\label{fig:3CT2regirregcomp}
		\includegraphics[scale=0.5]{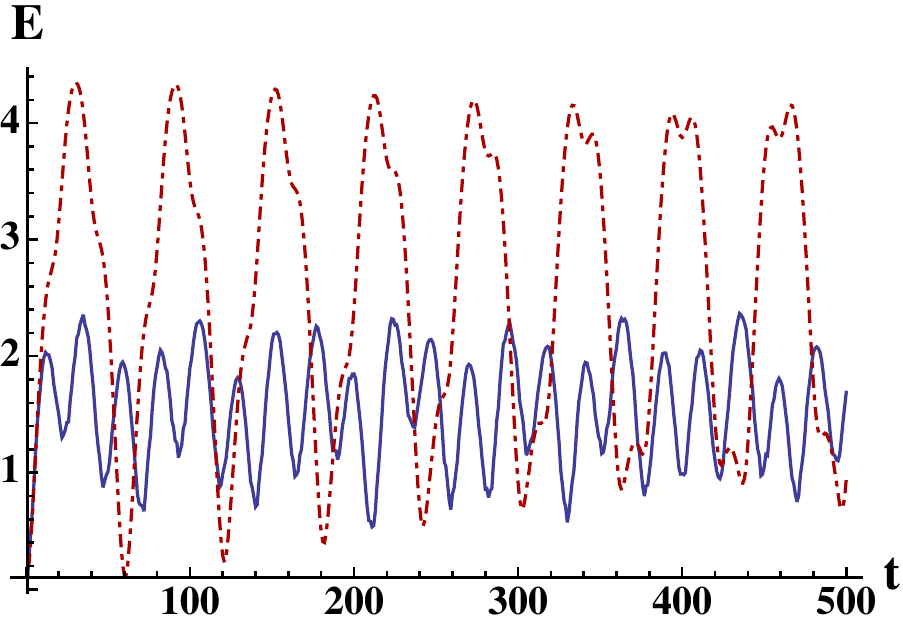}}
	\end{center}
  	\caption{Comparison of entanglement time series for regular graphs (blue solid lines) and their irregular counterparts (red dot-dashed lines), with an equal superposition initial state.}
  	\label{fig:regirregcomp}
\end{figure}

When we compare the time series for the irregular graphs with those for the regular graphs (Figure \ref{fig:regirregcomp}), we see that the modified $K_8$ graph has similar time series to the original $K_8$ graph. The time series appear to be similar in shape, with slightly different frequencies. On the other hand, the entanglement time series for a two-particle quantum walk on the modified $Q_3$ graph differs much more from its regular counterpart, with more complex behaviour occurring. This is also true for the unjoined Cayley tree, which has a more complex entanglement time series than the joined Cayley tree. 

\begin{figure}[htp]
    \newpage
	\begin{center}
	    \subfigure[Modified $K_8$ graph]{\label{fig:K8irrFeig}
		\includegraphics[scale=0.7]{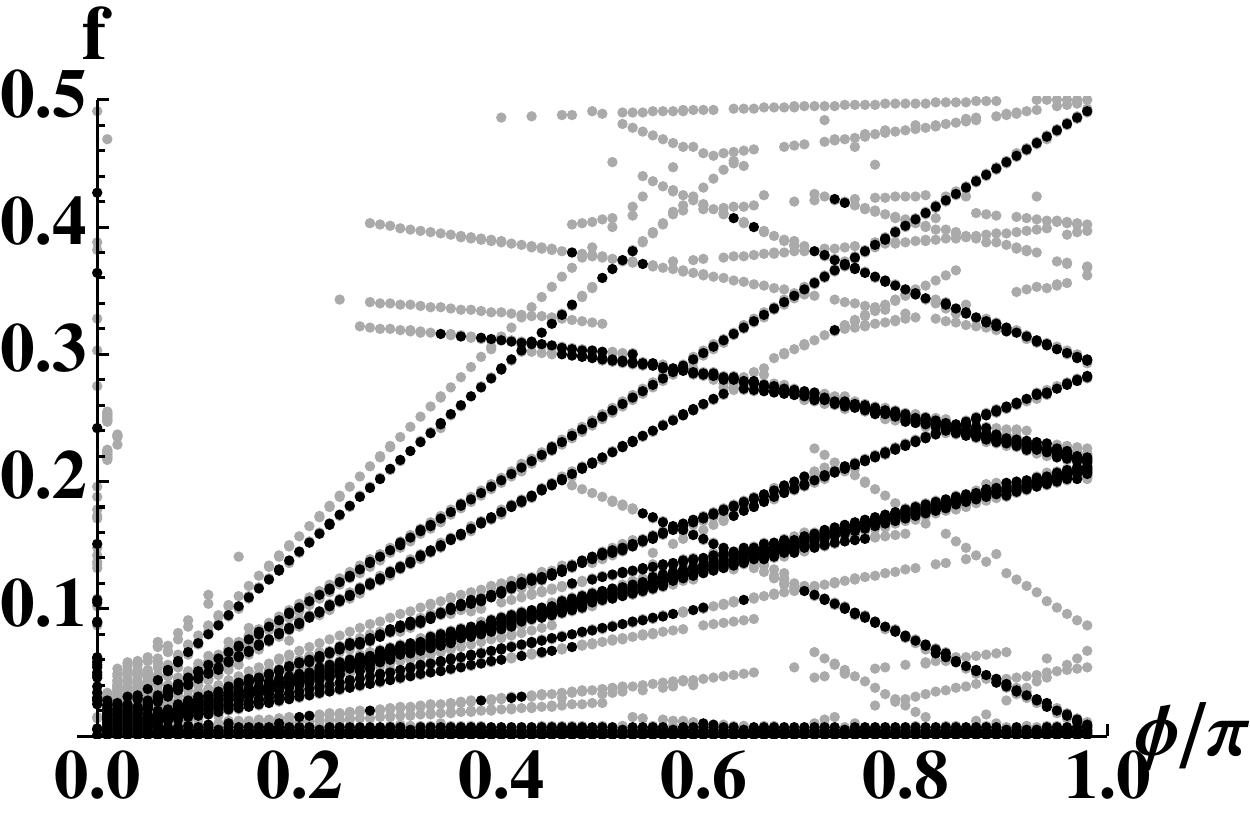}} \\
	    \subfigure[Modified $Q_3$ graph]{\label{fig:Q3irrFeig}
		\includegraphics[scale=0.7]{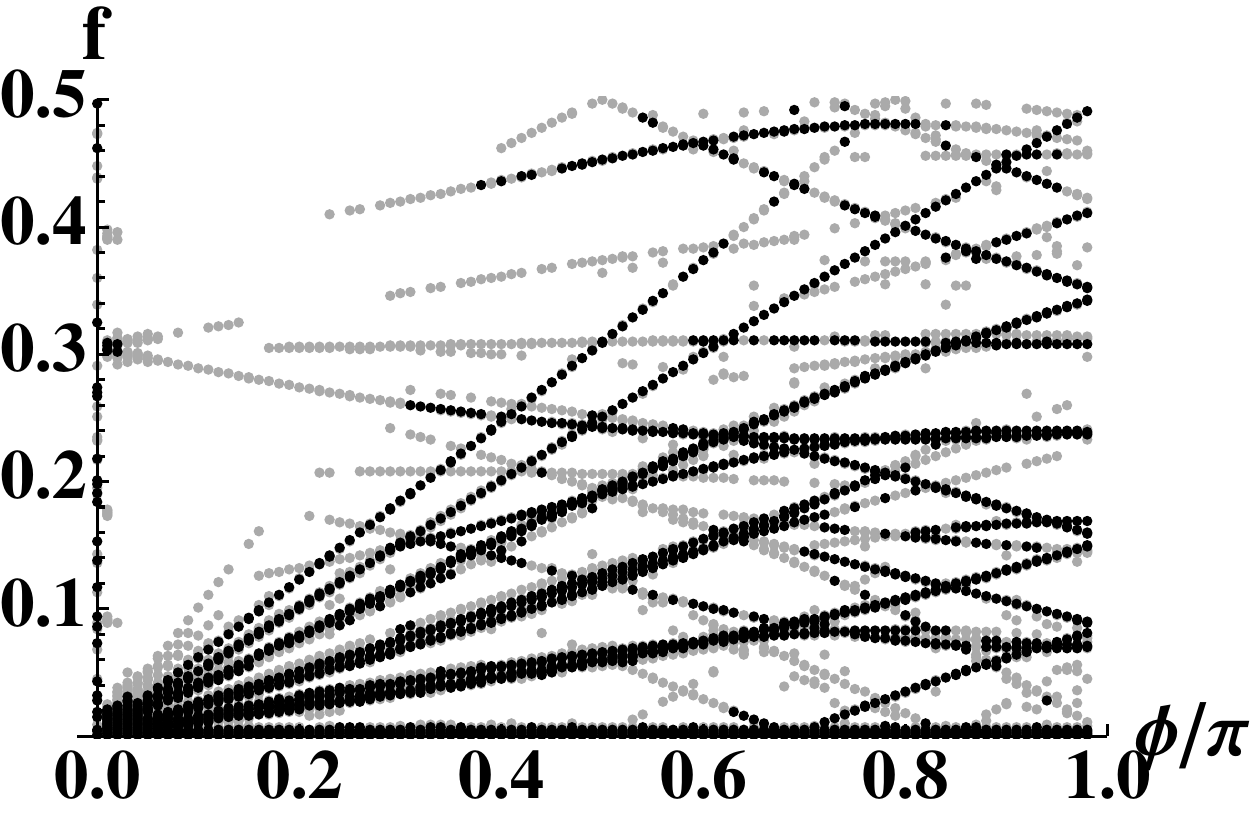}} \\
	    \subfigure[Unjoined 3CT2 graph]{\label{fig:3CT2Feig}
		\includegraphics[scale=0.7]{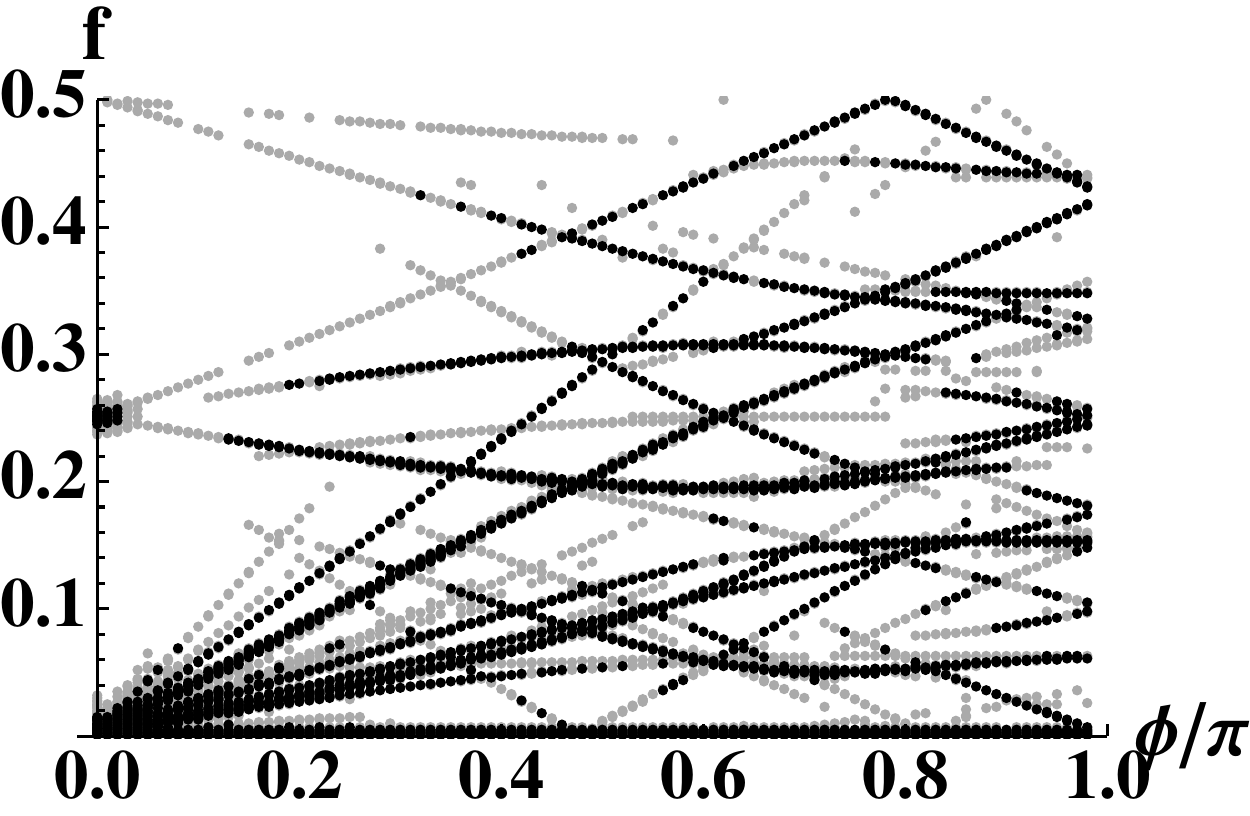}} \\
	\end{center}
	\caption{Feigenbaum diagrams of frequencies present in the entanglement
	time series of several irregular graphs under $\phi$-Grover interaction, with an equal superposition initial state.}
	\label{fig:irreg_Feig}
\end{figure}

This is more evident in the Feigenbaum diagrams for these time series, shown in Figure \ref{fig:irreg_Feig}, which depicts the frequencies present in the entanglement time series.  As expected, a reduction in complexity is also seen as $ \phi \rightarrow \pi $. The Feigenbaum diagrams for these graphs show that the entanglement time series for the irregular graphs are more complex than for the irregular graphs, with many more frequencies being present, and  nonlinear relationships between frequency and $\phi$ start to emerge (Figure \ref{fig:3CT2Feig}).

\begin{figure}[htp]
    \newpage
	\begin{center}
		\subfigure[Modified $K_8$ graph]{\label{fig:K8irrperturb}
		\includegraphics[scale=0.6]{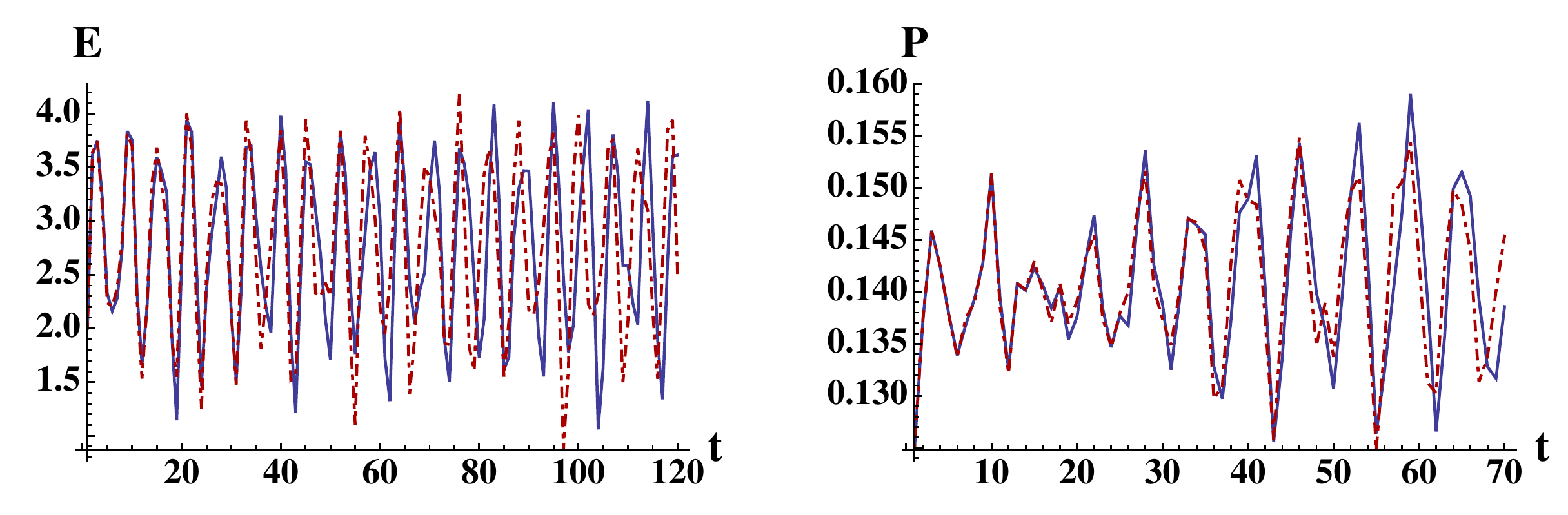}}
		\subfigure[Modified $Q_3$ graph]{\label{fig:Q3irrperturb}
		\includegraphics[scale=0.6]{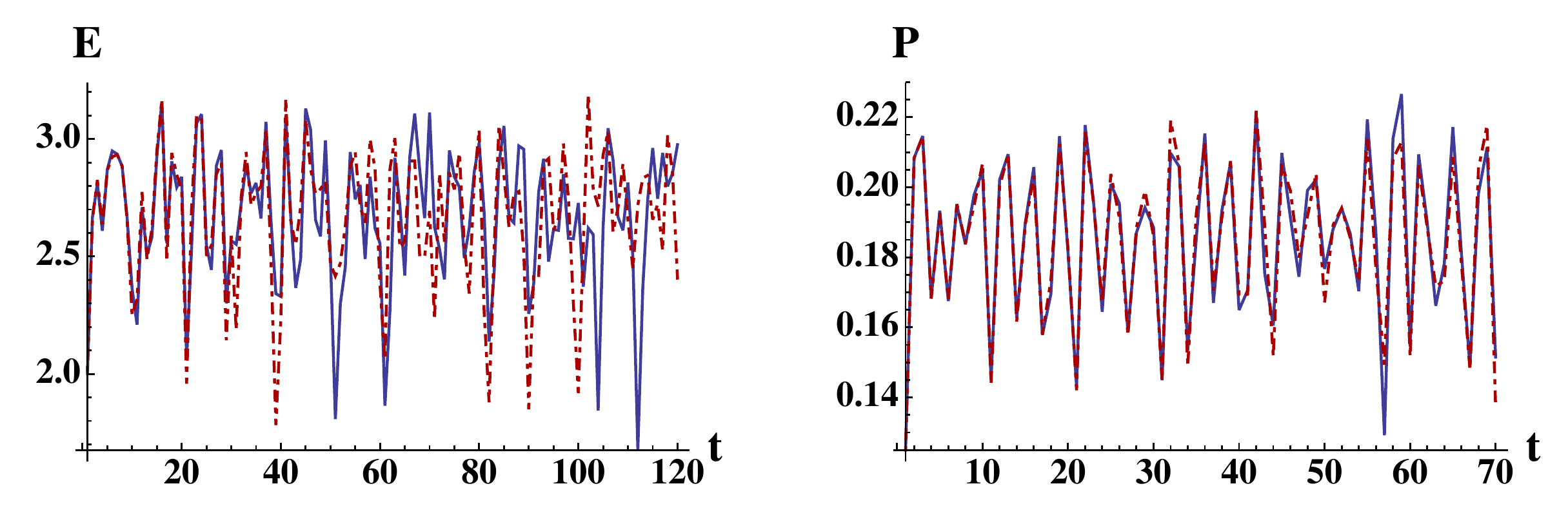}}
		\subfigure[Unjoined 3CT2 graph]{\label{fig:3CT2perturb}
		\includegraphics[scale=0.6]{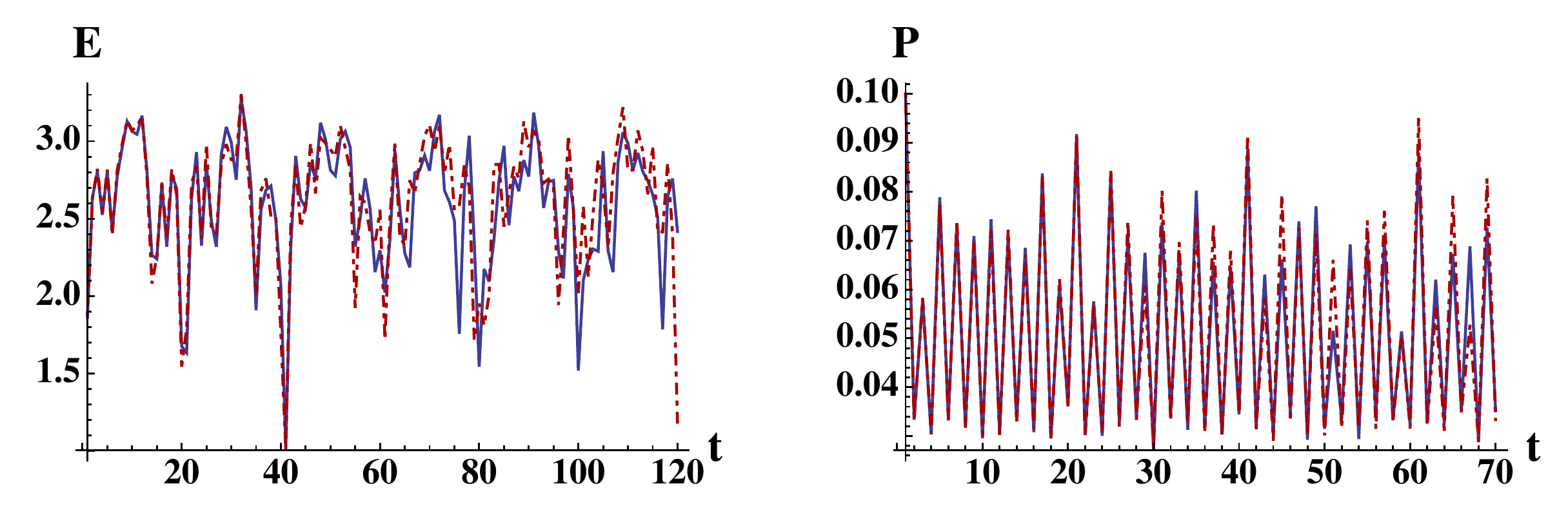}}
	\end{center}
	\caption{Time series of entanglement and single-particle probability, 
	for $ \phi = 0.75\pi $ (blue, solid) and $\phi = 0.76\pi $ (red, dashed) for a two-particle quantum walk under $ \phi$-Grover interaction beginning in an equal superposition initial state.}
	\label{fig:irreg_perturb}
\end{figure}

Like the regular graphs studied, the system also exhibits sensitivity to small perturbations in the $ \phi $ parameter for the irregular graphs, as shown in Figure \ref{fig:irreg_perturb}. It can be seen that the single-particle probabilities are less sensitive to these perturbations.  Again like the regular graphs studied, the initial state does not make a considerable difference in the frequency spectrum of the entanglement time series, although the shape of the time series does vary slightly, more so for these graphs than for the regular graphs.

\section{Conclusions}
\label{sec:conclusion}

In conclusion, we have demonstrated that while the probability time series for
simple degree-regular and degree-irregular graphs such as the $ K_8 $ graph and the 2nd generation 3-Cayley tree are generally stable regarding small changes to interaction strength, complex entanglement dynamics
can arise from a locally-interacting two-particle quantum walk, and these dynamics are sensitive to small perturbations to interaction strength.
Spectral analysis shows that the number of frequencies (and hence the
complexity) in the entanglement time series varies as a function of the
interaction strength $\phi$. The complexity in the corresponding Feigenbaum diagram seems to reflect the degree of symmetry in the underlying graphs, with the degree-irregular graphs showing much more complex Feigenbaum diagrams than the degree-regular graphs. 
The sensitivity of the entanglement to small changes in the strength of the particles' interaction could be a useful tool in probing small changes to systems and could be used in applications such as graph isomorphism.

\clearpage

%\bibliography{References}


\end{document}